\documentclass{iopjournal}

\usepackage{amssymb, amsmath, amsthm}
\usepackage{bm} 
\usepackage{mathrsfs}
\usepackage{physics}

\usepackage{bbold}
\usepackage{tensor}
\usepackage{caption}
\usepackage{siunitx}
\usepackage[many]{tcolorbox}
\usepackage{empheq}
\usepackage{subcaption}
\usepackage{float}
\usepackage{hyperref}
\usepackage[numbers,sort&compress]{natbib}
\usepackage{tcolorbox}
\usepackage[]{todonotes}
\usepackage{graphicx}
\usepackage{wrapfig}
\usepackage{mathtools} 
\usepackage{cleveref}
\usepackage{url}
\usepackage{algorithm}
\usepackage{booktabs}
\usepackage{algpseudocode} 
\usepackage{tabularx}
\usepackage{array}

\newcommand{\vc}{\mathcal{V}}

\newcommand*{\Reals}{\mathbb{R}}

\renewcommand{\[}{\begin{equation}\begin{aligned}}
\renewcommand{\]}{\end{aligned}\end{equation}}

\usetikzlibrary{patterns}

\usepackage{xcolor}
\usepackage{soul}

\renewcommand{\vc}{\mathcal{V}}

\begin{document}

\articletype{Paper} 
\title{Real-time virtual circuits for plasma shape control via neural network emulators}

\author{Alasdair Ross$^{1*\dagger}$, George K. Holt$^{2*\dagger}$, Kamran Pentland$^2$, Adriano Agnello$^1$, Nicola C. Amorisco$^2$, Pedro Cavestany$^1$, Aran Garrod$^1$, Timothy Nunn$^2$, Charles Vincent$^2$, Graham McArdle$^2$}

\affil{$^1$STFC Hartree Centre, Sci-Tech Daresbury, Keckwick Lane, Daresbury, Warrington, WA4 4AD, United Kingdom}

\affil{$^2$United Kingdom Atomic Energy Authority, Culham Campus, Abingdon, Oxfordshire, OX14 3DB, United Kingdom}

\affil{$^*$Authors to whom any correspondence should be addressed.}
\affil{$^\dagger$These authors contributed equally to this work.}

\email{alasdair.ross@stfc.ac.uk, george.holt@ukaea.uk}


\begin{abstract}
Reliable position and shape control in tokamak plasmas requires accurate real-time regulation of several strongly coupled shape parameters.
The control vectors that disentangle these couplings, referred to as \textit{virtual circuits} (VCs), enable independent shape parameter control for a specific Grad Shafranov (GS) equilibrium.
Numerical calculation of VCs is not currently feasible in real time, therefore VCs are usually computed prior to each experiment,  using a small number of reference GS equilibria sampled along the desired scenario trajectory, with each VC used to control the plasma within a preset time interval.
While effective near the reference equilibrium, this approach can lead to degraded performance as the plasma departs from the reference equilibrium and/or from the desired trajectory, and it complicates the design of robust control strategies for rapidly evolving plasma configurations. 
In this paper, we construct neural-network-based emulators of plasma shape parameters from which VCs can be derived, to provide the MAST Upgrade (MAST-U) plasma control system with state-aware VCs in real time.
To do this, we develop an extensive library of over a million simulated GS equilibria, covering a substantial portion of the MAST-U operational space. 
These emulators provide differentiable functions whose gradients can be rapidly computed, enabling the derivation of accurate VCs for real-time shape control. 
We perform extensive verification of the emulated VCs by testing whether they disentangle the control problem.
The neural-network-based approach delivers high accuracy and orthogonality across a diverse range of equilibria.
This work establishes the physical validity of emulated VCs as a scalable and general alternative to schedules of precomputed VCs. 

\end{abstract}

\section{Introduction}

Tokamaks require safe and reliable plasma position and shape controllers to reach high performance operating scenarios and advanced divertor configurations. 
To maintain the intended configurations throughout a discharge, the magnetic fields generated by the poloidal field (PF) coils are modulated on sub-millisecond timescales by the plasma control system (PCS) \cite{ariola2008magnetic,Walker_tok_plas_ctrl}. 
Rather than controlling the entire plasma separatrix, most tokamak plasma control systems target sets of globally defined shape parameters (e.g. X-points, midplane radii), gaps (i.e. distances between the separatrix and the first wall), or fixed poloidal isoflux locations \cite{Mele_2025}.
The control problem is inherently coupled: each PF coil influences the global magnetic equilibrium, and therefore affects all shape parameters simultaneously.
To manage this, linearised mappings -- commonly referred to as virtual circuits (VCs) -- are constructed to relate small changes in coil currents to variations in individual shape parameters around a reference equilibrium \citep{MCARDLE2020111764, wai2026tutorialinversionbasedshapecontrol}.
These mappings approximately decouple the system, enabling targeted and independent regulation of each shape parameter and supporting the design of modular control strategies.
This is the approach taken on MAST Upgrade (MAST-U), where real-time boundary shape control has been demonstrated for advanced divertor configurations \cite{Anand_2024}.

Multivariable shape regulation based on singular-value decomposition (SVD) was demonstrated at JET with the eXtreme Shape Controller, using combinations of shape errors and PF currents to balance tracking accuracy against control effort \cite{ariola2005xsc,ambrosino2008xsc}.
This framework was subsequently extended to simultaneous plasma shape and boundary flux control \cite{ariola2008flux}.

As VC calculation using numerical solvers does not have sufficiently low latency for real-time deployment, VCs are typically precomputed offline from a set of reference Grad–Shafranov (GS) equilibria distributed along the desired scenario trajectory, with each set applied over a predefined time interval. This approach is inherently local: as the plasma departs from the reference equilibrium, the accuracy of the linearisation degrades and the decoupling between shape parameters is progressively lost. As a result, unwanted coupling can arise, degrading controller performance and increasing demand on the PF coil set, thereby restricting access to certain plasma configurations. In addition, this workflow is highly manual, requiring expert intervention to select reference equilibria and tune control phases, which further complicates the design of robust controllers for rapidly evolving scenarios.

In this paper, we present an alternative approach in which neural networks (NNs) are developed that can be used to replace precomputed VC schedules within the control system.
Using a large dataset of synthetic MAST-U GS equilibria, we train an ensemble of NNs to predict several global shape parameters from input data spanning a wide portion of the MAST-U operating space (e.g. PF coil currents and plasma current density parameters).
The NNs yield differentiable functions from which VCs can be computed rapidly, making them suitable for real-time shape control.
This approach directly addresses the limitations outlined above while remaining a minimal extension to existing PCS architectures \citep{FPDT}.
It retains the familiar VC-based control framework, while replacing local linearisation with a dynamic, state-aware linearisation.
Crucially, it preserves the interpretability of control actions.
The methodology can be applied to other tokamaks using equilibrium datasets and shape descriptors appropriate to each device.

The intended control architecture consists of approximately decoupled feedback channels associated with selected shape descriptors.
Each channel acts through a combination of shaping coil currents, using the coupled response of the coil set.
We use seven MAST-U shape descriptors chosen for their operational relevance and availability for feedback control.
The question addressed here is whether emulator-derived VCs reproduce the physical response and approximate decoupling of GS-derived VCs within this established formulation.

The application of NNs for accelerated plasma shape control -- bypassing slow computations required to solve the GS equation -- is not itself new. 
Early work explored NNs for real-time plasma shape reconstruction \cite{bishop1992,bishop1995}, typically using relatively small models trained on limited datasets when compared with current standards. With modern computational resources, significantly more expressive models can now be trained over much broader regions of operating space.
More recently, machine learning methods have been applied across a range of tokamak modelling and control tasks.
NNs have been studied for GS equilibrium modelling and reconstruction \cite{wai2022,lao2022,PROKHOROV2020857}.
Recent approaches include convolutional physics-informed equilibrium and separatrix reconstruction \cite{bonotto2024planet}, multi-task reconstruction from magnetic measurements \cite{zheng2024efitnn}, and hybrid algorithms combining neural estimates with numerical equilibrium reconstruction \cite{zheng2025efitmini}.
A physics-informed TCV surrogate has been evaluated in real-time parallel operation and closed-loop simulations \cite{grandin2026surrogate}.
CARONTE instead fits physics-informed extreme learning machine output weights online to reconstruct the plasma boundary \cite{fiorenza2026caronte}.
The present study learns the forward mapping from equilibrium inputs to shape descriptors and validates the coil current sensitivities and VCs obtained from that mapping.
NNs have also been studied for vertical instability control and disruption avoidance \cite{rui2025,rasouli2013,2026NucFu..66b6012R, Orozoco_NN_RT_disruption,Tang-Deep_learning_disruption_prediction}.
Reinforcement learning methods \cite{RL_book_SuttonB98} have also been studied for active tokamak control \cite{DeTommasi2022,degrave2022,seo2024,tracey2024,kerboua2024,Char_offline_RL_tok_ctrl}.

The outline of this paper is as follows.
In \cref{sec:VC_intro}, we introduce VCs and their role in enabling plasma shape control on MAST-U.
In \cref{sec:dataset_generations}, we discuss the computational pipeline developed to construct the NN emulators, from the GS equilibrium library generation through to model training and performance.
In \cref{sec:results}, we evaluate the accuracy of the emulated VCs for different ensemble sizes and VC calculation methods (i.e. automatic differentiation vs. finite-differences), demonstrating the emulated VCs' effectiveness and quantifying their performance.
Finally, we conclude in \cref{sec:discussion} and outline concurrent follow-up work aiming towards deployment of AI-based shape controllers for MAST-U.

\section{Magnetic shape control} \label{sec:VC_intro}

In a tokamak, magnetic shape control is achieved by making adjustments to the PF coil voltages, thereby driving changes in currents in real time. 
Shape controllers monitor departures from predefined feedback reference waveforms on specific plasma shape parameters (e.g. midplane radii, X-point locations, or boundary gaps) or enforce requested feedforward drives \cite{FPDT}.
Here, we discuss how VCs map these departures or feedforward requests to the PF coil current adjustments required to enact the desired shape control.
We then explain this in the context of shape control on MAST-U to help motivate how we build emulators for these VCs that are suitable for real-time deployment. 

\subsection{Virtual circuits}
In the following, matrices are denoted by capital calligraphic fonts and vectors denoted with bold font. Let $\bm P \in \Reals^{n_p} $ be the vector of plasma shape parameters, $\bm I_\text{act} \in \Reals^{n_c}$ be the vector of active (controllable) PF coil currents (a subset of which are used for shape control and that we will refer to with $\bm I_\text{shape}$) and the central solenoid current, and $\bm \theta \in \Reals^{n_\theta}$ be a set of parameters that define the plasma current density profile.
The plasma equilibrium, and therefore the shapes $\bm{P}$, are fully defined by these parameters such that
\[ 
\label{eq:shape_function}
\bm P = \bm P(\bm I_\text{act}, I_p, \bm \theta).
\]
Plasma pressure is included in the GS equilibria underlying this mapping through the profile parameters $\bm\theta$.
For MAST-U, these include the ``Lao85'' pressure-profile coefficients described in \cref{sec:MAST-U}.

Given a small change in the coil currents used to control the plasma shape $\delta \bm I_\text{shape}$, the difference in plasma shape across the two corresponding Grad Shafranov equilibria is characterised by the \textit{sensitivity} or \textit{shape} matrix $\mathcal S$:
\begin{equation}
\label{eq:linearisation_jacobian}
\delta \bm P \ = \underbrace{\frac{\partial \bm P}{\partial\bm I_\text{shape}}}_{\mathcal{S}} \delta \bm I_\text{shape} + \mathcal{O}( \| \delta \bm I_\text{shape} \|^2) .
\end{equation}

During real-time shape control, changes in the plasma shape $\delta\bm{P}$ requested by a shape controller (whether in feedback or feedforward mode) are mapped to a corresponding change in the active coil currents by the VC matrix $\mathcal{V}$, such that:
\[
\label{eq:linear_inversion}
\delta \bm I_\text{shape} = \vc \delta \bm P.
\]
The VC matrix is the Moore--Penrose pseudoinverse of $\mathcal{S}$, computed by SVD, with each column representing the combination of PF coil current changes required to move a shape parameter by one unit.
This gives the coil current increment with minimum Euclidean norm among the least squares solutions while imposing no restriction on individual coil current limits.

As such, each VC approximately decouples the control of its associated shape parameter from the others, aiming to ensure that any changes to its target does not also affect the others \cite{wai2026tutorialinversionbasedshapecontrol}.

It can be seen in \cref{eq:linearisation_jacobian} that $\mathcal{V}$ is a function of the GS equilibrium from which it was computed, and therefore only valid in its proximity.
During experimental discharges, a small number of precomputed VCs are typically used, switching at preset times as the plasma response is expected to appreciably change. 
These VCs are generally built using finite difference methods, which requires solving multiple GS equilibria to calculate the response of the shape parameters to individual coil current perturbations.
This can take at least several seconds, making it unsuitable for real-time application.
Therefore, we instead build NNs to emulate \cref{eq:shape_function}, bypassing the need for multiple GS solves.
We then use these NNs to calculate sensitivity matrices $\mathcal{S},$ via finite differences or automatic differentiation, and therefore $\mathcal{V}$.
The latency of these operations makes this approach suitable for real-time deployment. 

\subsection{MAST-U specifics}\label{sec:MAST-U}

In this subsection, we discuss the inputs and outputs of \cref{eq:shape_function} and subsequent VC emulation in the context of the MAST-U tokamak, as listed in \cref{tab:shape_params}.
We emphasise that this subsection is machine-specific, but the framework and validation process presented in the rest of the paper are both machine-agnostic.

The set of active coils in MAST-U, whose currents are represented in $\bm I_\text{act}$, consists of a central solenoid, represented by $I_\text{solenoid}$, primarily used for plasma current drive, 10 up-down-symmetric coils for plasma shaping, represented in $\bm I_\text{shape}$, and one up-down-anti-symmetric coil for vertical stability control and positioning, represented by $I_\text{vert}$.
A complete description of the coil set can be found in \cite{MCARDLE2020111764}. 
To compute $\bm{P}$ in \cref{eq:shape_function}, the total plasma current $I_p$ and the plasma current density profile parameters $\bm{\theta}$ are also required. 
We use the ``Lao85'' \cite{Lao_1985} parametrisation for ${\bm \theta}$, which is routinely used in post-shot EFIT++ equilibrium reconstruction on MAST-U and is comprised of 4 free parameters: two characterising the internal pressure profile, $p'$, and two for the toroidal magnetic field profile, $FF'$. See \cite{pentland2024} for further details.

\begin{figure}[t!]
    \centering
    \begin{subfigure}{0.49\linewidth}
        \includegraphics[width=0.99\textwidth]{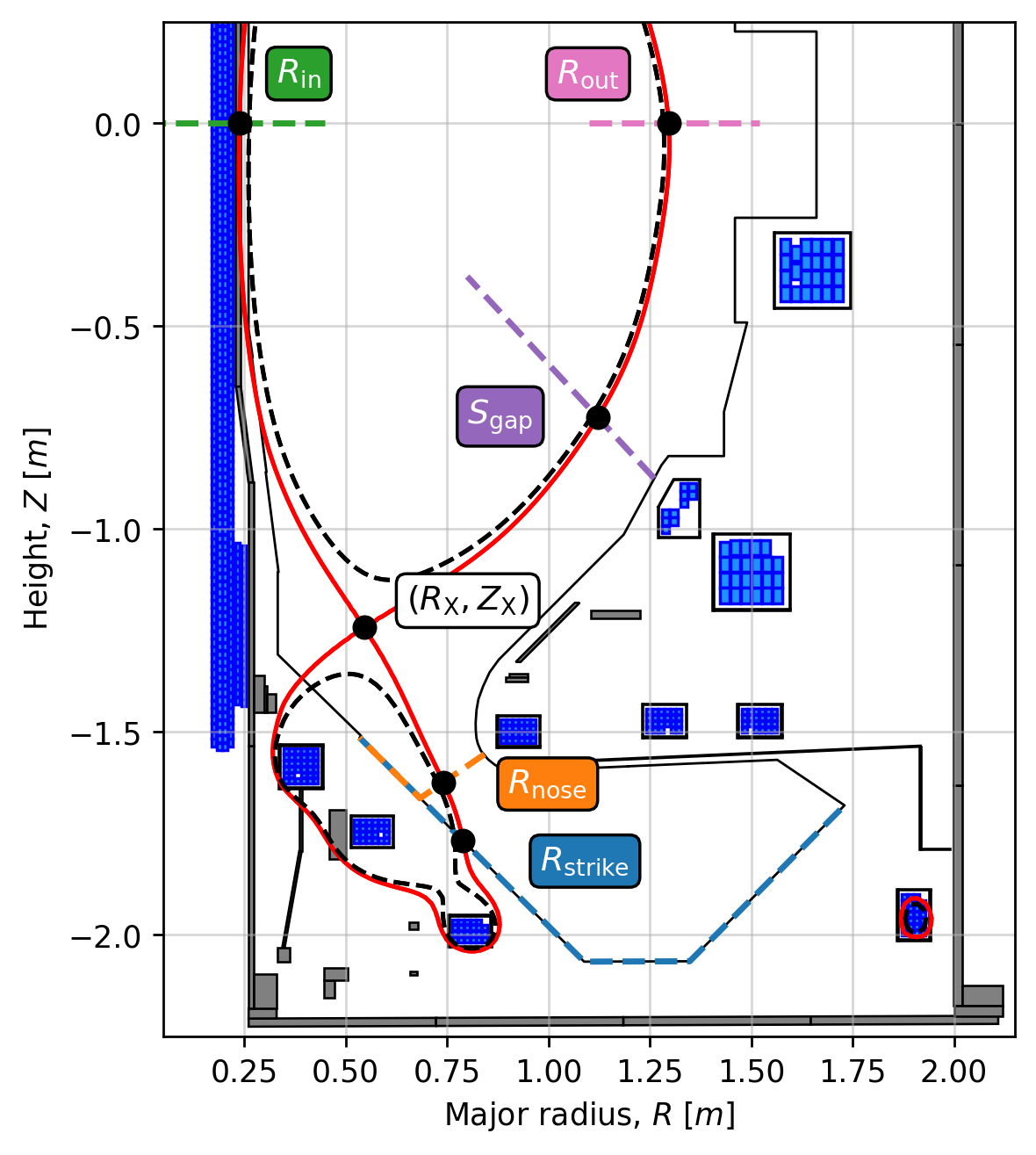}
        \caption{Limited plasma}
    \end{subfigure}
    \begin{subfigure}{0.49\linewidth}
        \includegraphics[width=0.99\textwidth]{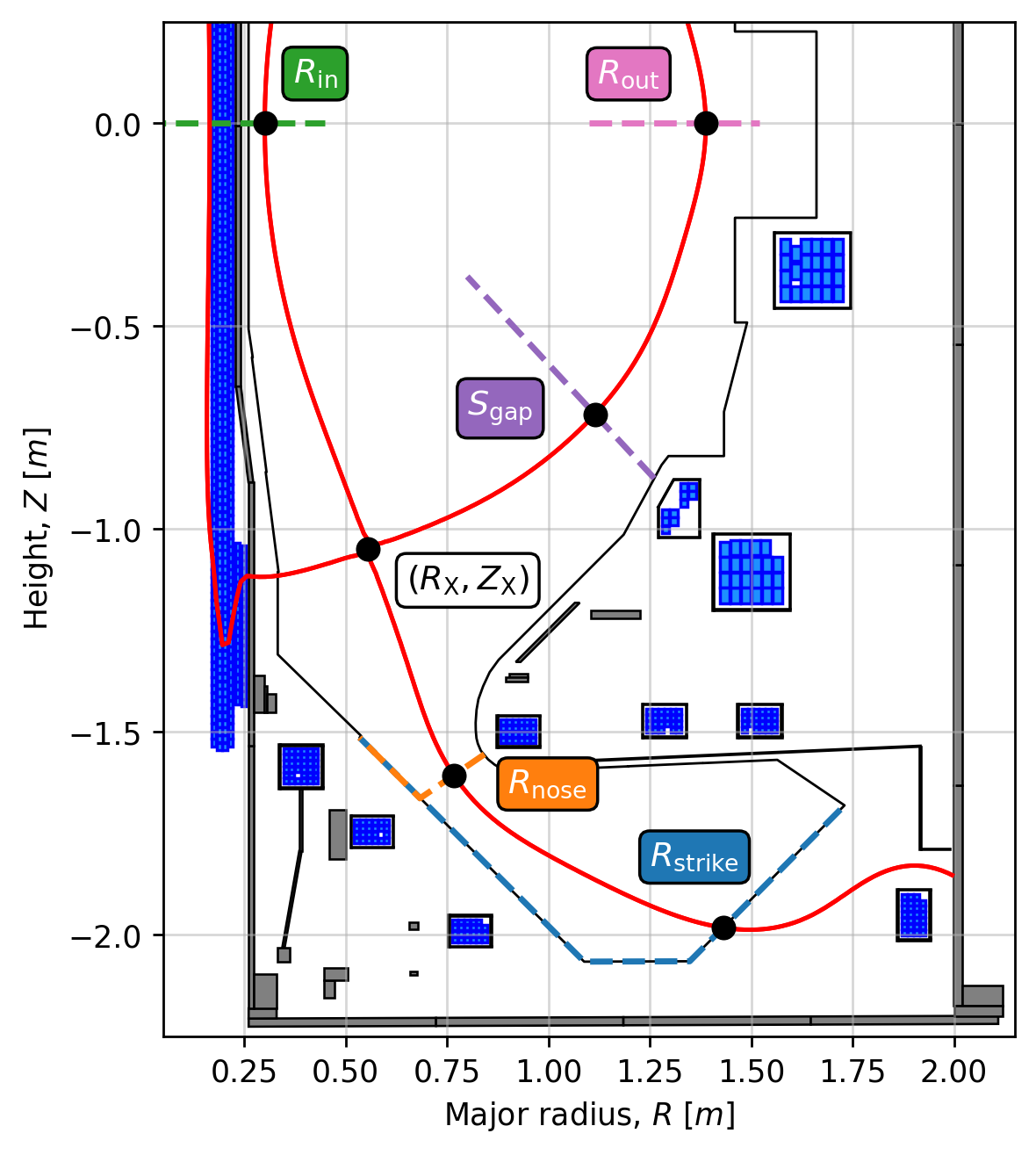}
        \caption{Diverted plasma}
    \end{subfigure}
    \caption{
    Poloidal cross-section of the lower half of MAST-U with the PXPS (solid red) for limited and diverted equilibria from our library.
    Coloured dashed line segments indicate the lines on which the shape parameters listed in \cref{tab:shape_params} are defined by intersection with the PXPS.
    In the limited case (a), the LCFS is shown as a dashed black line, distinct from the PXPS. In the diverted case (b), the PXPS coincides exactly with the LCFS.
    Also shown are the PF coils in solid blue, passive conducting structures in grey, and first wall in solid black.
    }
    \label{fig:MAST_U-shapes}
\end{figure}

We focus on a set of seven MAST-U plasma shape parameters that are desirable to and available for feedback control in MAST-U discharges as reconstructed in real time \cite{kochan2023, Anand_2024}.
These are listed in \cref{tab:shape_params} and illustrated in \cref{fig:MAST_U-shapes} by black markers.
The sensitivity matrix therefore has dimensions $7 \times 10$.
The actuator count alone does not establish controllability, which also depends on the singular values and the available coil current headroom.
Each shape parameter is defined by the intersection of a different virtual line with the primary X-point separatrix\footnote{The PXPS is defined as the poloidal flux surface passing through the primary X-point.} (PXPS, shown in solid red).
These virtual lines are shown with dashed style in \cref{fig:MAST_U-shapes} with different colours. 
For diverted equilibria (such as that shown in the right-hand panel), the PXPS and the last closed flux surface (LCFS) are identical. 
This is not the case for limited equilibria. 
However, we explicitly choose to use the PXPS rather than the LCFS for limited equilibria as this allows emulators to provide non-trivial derivatives $\mathcal{S}$ in this regime, which are necessary for meaningful control. 
This can be clearly seen in the left-hand panel for the inboard midplane radius, which our definition places outside the first wall. This allows the emulators to produce VCs that are able, for example, to push the plasma off the centre column when required.
Similarly, for $R_{\text{nose}}$, we extend the intersection contour onto the divertor tiles so that VCs for $R_{\text{nose}}$ are available even when the divertor leg does not cross into the divertor chamber. 

In previous work, it was observed that emulation of $R_\text{strike}$ is particularly challenging \cite{Emulated_VC_IEEE}.
This is a consequence of the difficulty posed by robust $R_\text{strike}$ characterization in the training library, especially for uncommon equilibrium configurations.
To ameliorate this, we ensure a robust calculation of $R_\text{strike}$ by computing the quantity using three distinct approaches:
{\it i}) intersection of the PXPS with the lower divertor tiles (shown in dashed blue in \cref{fig:MAST_U-shapes}),
{\it ii}) tracing the flux line from a point slightly outboard of the primary X-point until it intersects with the vessel wall, and
{\it iii}) tracing the flux line from a point slightly outboard of $R_\text{out}$ until it intersects with the vessel wall.

\begin{table}[t!]
\centering
\caption{MAST-U-specific definitions of the input and output parameters in \eqref{eq:shape_function}. Refer to \cref{fig:MAST_U-shapes} for illustrative definitions of the shape (output) parameters.}
\label{tab:shape_params}
\begin{tabular}{|l|l|l|}
\hline
Inputs & Description & Units \\
\hline \hline
$\bm I_\text{shape}$ & Vector of PF coil currents used for shape control ($n_c = 10$) & [\SI{}{\ampere}] \\
$I_\text{solenoid}$ & Solenoid current & [\SI{}{\ampere}] \\
$I_\text{vert}$ & PF coil used for vertical stability and placement control & [\si{\ampere}] \\
$I_\text{p}$ & Total plasma current & [\SI{}{\ampere}] \\
$\alpha_0,\alpha_1$ & Pressure coefficients in ``Lao85'' plasma current density profile \cite{Lao_1985} & [\si{\pascal/\weber}] \\
$\beta_0,\beta_1$ & Toroidal-field coefficients in ``Lao85'' plasma current density profile  \cite{Lao_1985} & [\si{\tesla}] \\
\hline \hline
Outputs & &  \\
\hline \hline
$R_\text{in}$ & Inboard midplane radius & [\SI{}{\meter}] \\
$R_\text{out}$ & Outboard midplane radius & [\SI{}{\meter}]\\ 
$R_\text{X}$ & Radial position of the lower X-point & [\SI{}{\meter}]\\
$Z_\text{X}$ & Vertical position of the lower X-point & [\SI{}{\meter}]\\
$R_\text{strike}$ & Radial position of the lower outboard strike point & [\SI{}{\meter}]\\
$R_\text{nose}$ & Radial position of the separatrix entering the lower divertor chamber & [\SI{}{\meter}]\\
$S_\text{gap}$ & Squareness gap between the core plasma and first wall & [\SI{}{\meter}] \\
\hline

\hline 

\end{tabular}
\end{table}

\section{Equilibrium library generation, model training and performance} \label{sec:dataset_generations}
With the shape parameters and input space defined, we now describe the framework developed to generate the training data (i.e., a library of GS equilibria), train the NN emulators, and validate their performance.

\subsection{Neural network motivation} \label{sec:data_model_generation}

Feedforward NNs are universal approximators: under certain constraints, they can approximate continuous or $L_p$-integrable functions to arbitrary accuracy \cite{Cybenko1989, HORNIK1989, LESHNO1993}.
This approximation capability extends to the derivatives of the target function under sufficient smoothness conditions of the activation functions \cite{HORNIK1990, HORNIK1991}. 

Evaluation of a trained NN can be rapid, enabling low-latency inference with throughput scaling as a function of model complexity.
This provides the prospect that NNs could be used to perform real-time calculations in a plasma control system.

We train feedforward NNs to predict the shape parameters, emulating \cref{eq:shape_function}.
We then use these emulators to evaluate the derivatives that comprise the sensitivity matrix in \cref{eq:linearisation_jacobian}.
Alternatively, one could train NNs to emulate the Jacobians themselves.
We explicitly do not choose this route.
With significantly more outputs, this would require a much larger model, and in turn a larger dataset for training.
The dataset would also require the Jacobians, which would incur significantly greater computational cost for each equilibrium in the dataset.

\subsection{Equilibrium library generation} \label{sec:dataset_generation}

Training a robust NN for tokamak plasma control requires a large dataset containing the inputs (currents and profile parameters) and outputs (shape parameters), which are extracted from a library of GS equilibria.
We opt to use synthetic GS equilibria provided by the free-boundary equilibrium code FreeGSNKE \cite{freegsnke}.
This offers two distinct advantages over relying solely on prior plasma discharge data.
Firstly, the volume of historic data from previous MAST-U discharges is likely insufficient to train an accurate and robust emulator. 
Secondly, emulators trained solely on historic data would be valid only in the regime of previously observed plasma configurations and shapes, whereas a purpose of this work is to develop the capability of controlling as-yet-unseen configurations not yet demonstrated in experiments.

We generate such a dataset using a Markov Chain Monte Carlo (MCMC) approach to explore the input space, and use initial ``seed'' equilibria from historic plasma discharges as starting points for the Markov chains.
Full details of the MCMC algorithm used can be found in \cite{Agnello24}.
The MCMC consists of random walks that preferentially sample equilibria with desirable shapes and coil current configurations.
By construction, the MCMC sampler does not entirely prevent undesirable configurations from appearing in the dataset, but it steers the sampling away from uninteresting regions of parameter space, e.g., by penalising equilibria with displaced magnetic axes or excessively short connection lengths\footnote{The connection length is the distance measured along a magnetic field line in the scrape-off layer from the outboard midplane to the divertor target.}, or equilibria with coil currents that are beyond machine operating limits.
This approach is advantageous to emulator design for the purpose of control, as the desired and expected regions of parameter space are densely sampled, while other regions are still represented.

A single starting equilibrium was used by \cite{Agnello24} to initialise the chains. Here, in order to explore a much larger region of parameter space, we use multiple equilibria from historic MAST-U shots as seeds for the MCMC walkers. A total of \num{3999} seed equilibria were used, sampled uniformly from the flat-top phase of \num{33} shots that were chosen to cover a wide range of MAST-U plasma configurations. See \cref{app:seeding_shot_descriptions} for details of the shots used. 

\begin{figure}[ht]
    \centering
    \includegraphics[width=1\linewidth]{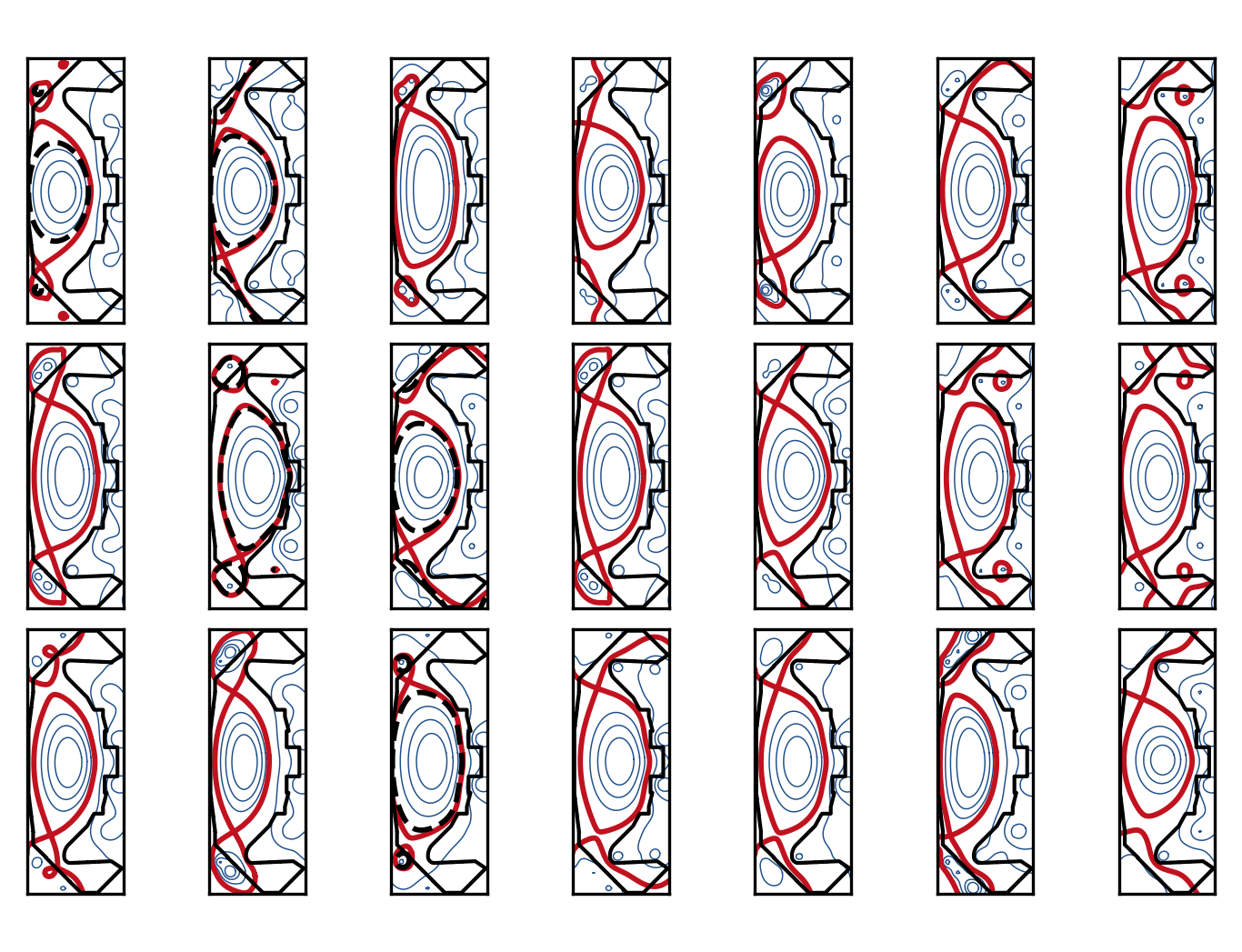}
    \caption{Representative MAST-U equilibria from the synthetic dataset generated by the MCMC walker algorithm.
    In each panel, the primary X-point separatrix (PXPS) is shown as a red solid line.
    The last closed flux surface (LCFS) is shown as a dashed black line for limited equilibria.
    Also shown are contours of poloidal flux in blue and the first wall in solid black.
    Together, these examples illustrate the diversity of plasma shapes and divertor configurations present in the dataset.}
    \label{fig:equilibria}
\end{figure}

The MCMC exploration of the input space generated \num{1676423} simulated MAST-U equilibria, including both limited and diverted plasmas, covering a variety of plasma shapes, positions, and divertor leg configurations.
To illustrate the diversity of equilibria contained in the dataset, \cref{fig:equilibria} shows a representative selection of plasma cross-sections drawn from the synthetic library.
These examples span a range of core shapes and divertor geometries, including both limited and diverted configurations.

\begin{figure}[h!t]
    \centering
    \includegraphics[width=0.9\linewidth]{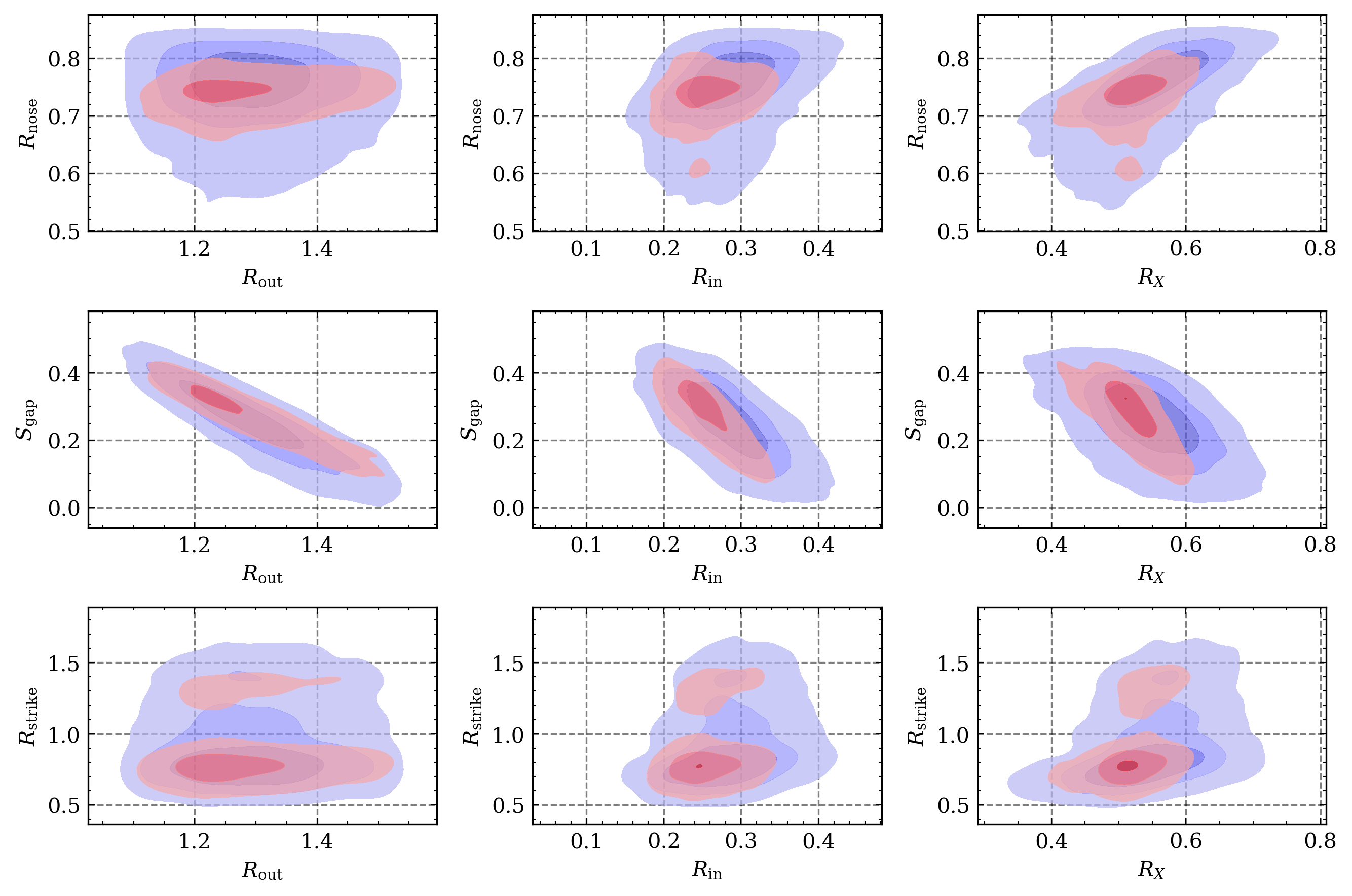}
    \caption{Kernel density estimation (KDE) plots for a selection of shape parameters for a sample of \num{10000} equilibria, showing MCMC starting seeds (red) and MCMC chain equilibria (blue). 
    }
    \label{fig:swarm_plot}
\end{figure}

To further quantify the span of equilibria generated by the MCMC algorithm compared to the seed equilibria, kernel density estimation (KDE) plots for a variety of shape parameters, showing both the seed and synthetically generated equilibria, are shown in \cref{fig:swarm_plot}.
The blue region, corresponding to the synthetically generated equilibria, spans a larger region of parameter space than the red region, which corresponds to the initial seed equilibria from real historic shots.
The seed distributions also show multimodality.
This is particularly noticeable in $R_\text{strike}$, for example, where a seed shot may have either a conventional-divertor or a super-X-divertor configuration~\cite{Harrison_2024_SuperX}.
The synthetic equilibria populate the gaps between these configurations, which are experimentally realised during the divertor leg sweep.

To address the issue with $R_\text{strike}$ discussed in \cref{sec:MAST-U}, we filter the dataset to remove uncertain strike-point values.
If the radial coordinates found by the three methods agree to within a tolerance of \SI{3}{\centi\metre}, the example is kept in the dataset; otherwise, it is discarded.
This filtering yields a final cleaned and filtered dataset consisting of \num{918124} equilibria.
All equilibria in this data set satisfied the same relative convergence tolerance of $10^{-5}$ in the FreeGSNKE forward static solve.
The convergence measure is the range (maximum minus minimum) of the nonlinear poloidal flux residual divided by the range of the plasma poloidal flux over the computational grid.
Supplementary diagnostics of the synthetic dataset, including marginal distributions of model inputs (\cref{fig:input_feature_distributions}) and outputs (\cref{fig:output_feature_distributions}), two-dimensional input--output coverage plots (\cref{fig:input_output_alpha_grid}), and pairwise output correlations (\cref{fig:pearson_output}), are provided in \cref{sec:synthetic_data_distributions}.

\subsection{Neural network training}
We train feedforward NNs to emulate the shape parameters $\bm{P} \in \Reals^7$, given the input vector
\[
\bm x = \left( \bm I_{\mathrm{act}}, I_p, \bm \theta \right) \in \mathbb{R}^{17},
\]
where $\bm{x}$ contains the active PF coil currents, total plasma current, and profile parameters, and the target $\bm{P}$ is the corresponding vector of seven shape parameters.
The network output is
\[
\hat{\bm P} = f_{\bm \phi}(\bm x) \in \mathbb{R}^7,
\]
where \(f_{\bm \phi}\) denotes the NN with trainable parameters \(\bm \phi\). 
The training objective is to learn an approximation
\[
f_{\bm \phi}(\bm x) \approx \bm P(\bm I_{\mathrm{act}}, I_p, \bm \theta).
\]

All models used in this work are fully connected multilayer perceptrons with a linear output layer, trained for supervised multi-output regression.
Prior to training, the full dataset was divided into disjoint training and test subsets using an 80/20 split.
The training set was then partitioned into \(K=5\) folds for cross validation.
This splitting yields a test set with \num{183625} equilibria, training folds of \num{587599} equilibria and validation folds of \num{146900} equilibria.
To improve optimisation and place all variables on comparable scales, both inputs and targets were standardised within each fold using statistics computed from the training partition of that fold only:
\[
\label{eq:normalisation}
\tilde{x}_j = \frac{x_j-\mu_j^{(x)}}{\sigma_j^{(x)}},
\qquad
\tilde{P}_k = \frac{P_k-\mu_k^{(P)}}{\sigma_k^{(P)}}.
\]
Here, $x_j$ denotes the $j$th component of the input vector, $P_k$ denotes the $k$th component of the shape parameter vector, and $\mu$ and $\sigma$ denote the corresponding component-wise means and standard deviations computed from the training partition of the fold.
The indices run over $j=1,\ldots,17$ input components and $k=1,\ldots,7$ output components.
The same transformations were then applied to the corresponding validation data.

For a given fold, model parameters were fit by minimising the mean-squared error (MSE) on the scaled targets,
\[
\mathcal{L}(\bm \phi)
=
\frac{1}{N_{\mathrm{train}}}
\sum_{i=1}^{N_{\mathrm{train}}}
\left\|
f_{\bm \phi}(\tilde{\bm x}^{(i)}) - \tilde{\bm P}^{(i)}
\right\|_2^2.
\]
Here, $i$ indexes the training examples in the fold and $N_{\mathrm{train}}$ is the number of such examples.

Hyperparameters were selected using a distributed random search implemented with Optuna \cite{akiba2019optuna}.
The search space included network width, network depth, activation function, optimiser, kernel initialiser, batch size, and initial learning rate.
For each sampled hyperparameter configuration \(\lambda\), a separate NN was trained on each cross-validation fold, and the configuration was scored using the mean of the best validation losses across folds,
\[
J(\lambda)
=
\frac{1}{K}
\sum_{m=1}^{K}
\mathrm{MSE}^{(m)}_{\mathrm{val},\mathrm{best}}(\lambda),
\]
where \(\mathrm{MSE}^{(m)}_{\mathrm{val},\mathrm{best}}\) denotes the minimum validation MSE attained during training on fold \(m\).
This provides a held-out-fold objective for selecting each candidate hyperparameter configuration.

Training used early stopping on validation loss, with restoration of the best weights, together with automatic reduction of the learning rate when the validation loss plateaued.
During the hyperparameter search, unpromising trials were additionally pruned using intermediate validation loss values from the first cross-validation fold, allowing computational effort to be concentrated on more promising regions of hyperparameter space.
The search was terminated when either no further improvement was obtained after a sufficient number of completed trials or the allocated wall-clock budget was reached.

After completion of the hyperparameter search, the top \(N=8\) completed trials, ranked by cross-validated mean validation MSE, were retrained to produce the final surrogate models used in this work.
Each of these models was trained on the training portion of the first fold of \num{587599} equilibria, using the corresponding validation subset for early stopping.
We use the first fold for this final retraining because the validation subset is used for early stopping as a form of regularisation.
Consequently, this subset must remain held out from the weight updates of the final models, and the final models are not retrained on the full training set.
The final predictor is an ensemble average using the simple arithmetic mean,
\[
\label{eq:ensemble_model}
\hat{\bm P}_{\mathrm{ens}}(\bm x)
=
\frac{1}{N}
\sum_{r=1}^{N}
f_{\bm \phi_r}(\bm x).
\]
Unless otherwise stated, this ensemble predictor is used throughout the remainder of the paper.

The sensitivity matrix derived from the emulators can be obtained directly from the ensemble model as
\[
\hat{\mathcal S}
=
\frac{\partial \hat{\bm P}_{\mathrm{ens}}}{\partial \bm I_{\mathrm{shape}}}
=
\frac{1}{N}
\sum_{r=1}^{N}
\frac{\partial f_{\bm \phi_r}}{\partial \bm I_{\mathrm{shape}}},
\]
which provides the learned approximation to the shape response Jacobian introduced in \cref{eq:linearisation_jacobian}. This can be obtained either directly by automatic differentiation of the ensemble model, or using a finite difference method. We compare both approaches in later sections.

\subsection{Model performance} \label{sec:model_performance}
In this subsection we present evaluations of the forward prediction capability of the ensemble of trained models in \cref{eq:ensemble_model}. 
A table of the architectures and training hyperparameters for the models in the ensemble is provided in \cref{app:model_architectures}.
Evaluation metrics, including MSE, mean absolute error (MAE), and root mean squared error (RMSE), for the ensemble of models evaluated on the test dataset are shown in \cref{tab:model_performace_test_both}.
These metrics are provided for both dimensionless data (scaled using the first-fold training data according to \cref{eq:normalisation}) and physical (unscaled) data in units of metres.
The forward prediction performance demonstrated by these models is high, with accurate values and low variance and bias.

The results reported in the following sections show that the ensemble model is sufficiently well-trained to be used for shape control, demonstrating that the metrics reported in \cref{tab:model_performace_test_both} are a good indicator of the efficacy of a downstream control policy based on these NNs.

We provide a graphical summary of the model predictive performance in \cref{app:model_prediction_performance}.

\begin{table}[h!t]
    \centering
    \caption{Model performance metrics -- mean squared error (MSE), mean absolute error (MAE), and root mean squared error (RMSE) -- evaluated on the test set for the ensemble of models.}
    \label{tab:model_performace_test_both}
    \begin{subtable}{\linewidth}
        \label{tab:ensemble_metrics_unscaled_test}
        \centering
        \caption{Metrics computed in physical units, following inverse transformation of both predictions and targets, quantifying the residual systematic uncertainties from the emulator models.}
        \begin{tabular}{l|rrrrrrr}
        \toprule
         & $R_\text{in}$ & $R_\text{out}$& $Z_\text{X}$ & $R_\text{X}$ & $R_\text{strike}$ & $R_\text{nose}$ & $S_{\text{gap}}$ \\
        \midrule
        \textbf{MSE} $(m^2)$ & 8.13e-06 & 3.87e-05 & 6.55e-06 & 3.55e-06 & 1.76e-04 & 5.43e-07 & 2.80e-05 \\
        \textbf{MAE  $(m)$} & 5.50e-04 & 1.35e-03 & 5.97e-04 & 5.05e-04 & 2.40e-03 & 3.23e-04 & 1.10e-03 \\
        \textbf{RMSE $(m)$ } & 2.85e-03 & 6.22e-03 & 2.56e-03 & 1.88e-03 & 1.33e-02 & 7.37e-04 & 5.29e-03 \\
        \bottomrule
        \end{tabular}
    \end{subtable}

    \vspace{1em}

    \begin{subtable}{\linewidth}
        \centering
        \label{tab:ensemble_metrics_scaled_test}
        \caption{Metrics computed in standardised units, where the ground-truth data have been normalised to have zero mean and unit variance, to highlight the relative residual scatter around the model predictions.}
        \begin{tabular}{l|rrrrrrr}
        \toprule
         & $R_\text{in}$ & $R_\text{out}$& $Z_\text{X}$ & $R_\text{X}$ & $R_\text{strike}$ & $R_\text{nose}$ & $S_\text{gap}$ \\
        \midrule
        \textbf{MSE } & 3.26e-03 & 4.45e-03 & 9.91e-04 & 8.30e-04 & 2.89e-03 & 1.76e-04 & 3.09e-03 \\
        \textbf{MAE } & 1.10e-02 & 1.45e-02 & 7.35e-03 & 7.72e-03 & 9.74e-03 & 5.82e-03 & 1.15e-02 \\
        \textbf{RMSE } & 5.71e-02 & 6.67e-02 & 3.15e-02 & 2.88e-02 & 5.38e-02 & 1.33e-02 & 5.56e-02 \\
        \bottomrule
        \end{tabular}
        
    \end{subtable}
\end{table}

A C++ implementation for real-time VC computation is described in the accompanying integration study \cite{RTVC} (Sections 3 and 4).
For the ensemble of models, server requests on development hardware comparable to the MAST-U PCS have a mean round-trip time of \SI{0.59}{\milli\second} and a 99th percentile of \SI{0.93}{\milli\second}, measured from sending inputs to retrieving outputs.
The integrated implementation, using finite difference Jacobians and subsequent pseudoinversion, generates updated VCs on a \SIrange{5}{5.6}{\milli\second} cycle during shot replay tests.

\section{Validation of VCs derived from the emulators} \label{sec:results}
In this section, we evaluate the performance of the VCs derived from the NN emulators.
The key requirement is that these VCs reproduce the defining properties of physically derived VCs: they must produce shape parameter shifts of the correct magnitude while maintaining approximate orthogonality between parameters.
In other words, each VC should induce the requested change in its target parameter without significantly affecting the others.

To assess this, we follow a similar procedure as in \cite{Emulated_VC_IEEE}.
We apply VC-derived coil current perturbations to a set of equilibria and evaluate the resulting plasma response using FreeGSNKE.
This allows us to directly test whether the emulated VCs correctly capture the local shape sensitivities.
We further compare different approaches to computing the VCs from the emulators, including automatic differentiation and finite differences, as well as benchmarking against VCs obtained from numerical GS solutions.
As this validation only involves static GS equilibria, we refer to this as {\it static} validation.
A companion work \cite{dynamic_val_paper}
explores using the obtained VCs in closed-loop evolutive equilibrium simulations, providing {\it dynamic} validation results.

We evaluate the static performance of the emulator-based VCs on a set of \num{5000} equilibria taken from the test dataset.
We focus on configurations that have operational relevance for MAST-U, which we enforce by limiting this set to only include equilibria generated early in the chain of parameter exploration.
As each chain seeded by a MAST-U equilibrium previously obtained in experiment was run for 300 MCMC steps, here we use equilibria with iteration number less than 50.
We also exclude limited equilibria, since on MAST-U these would only rarely be the subject of active shape control and, in such cases, one would not in practice seek to control $R_\text{nose}$, $R_\text{strike}$, $R_X$ or $Z_X$.
Omitting them therefore yields a more machine-relevant evaluation of VC performance. 

For each equilibrium, a VC is computed with both the emulators and with FreeGSNKE.
Each VC is then used to apply a \SI{5}{\milli\metre} requested shift ($ \delta \bm P_{\text{req}}$) in each shape parameter, one at a time.
In both cases, the resulting plasma response, in the form of shifted shape parameters, is obtained by solving the GS problem using FreeGSNKE with the updated coil currents obtained from the respective VCs.
The shape parameters are then recomputed using the new equilibrium:
\[ \label{eq:requested_shift}
\delta \bm P |_{GS} = \bm P(\bm I_{\text{act}} + \mathcal V \delta \bm P_{\text{req}}, I_p, \bm \theta) - \bm P( \bm I_{\text{act}}, I_p, \bm \theta).
\]
We refer to the post-VC shift $\delta \bm P |_{GS}$ as the \textit{realised shift}.
The requested shift size $|\delta \bm P_\text{req}| = \SI{5}{\milli\metre}$ is chosen to be sufficiently small that it lies in the linear regime where the VC is likely to be accurate, while also being significantly larger than numerical precision on the shape parameters themselves.
Larger requested shifts are expected to produce increasing departures from the local linear approximation, leading to larger errors in the realised target shifts and greater residual coupling between shape parameters for both GS-derived and emulator-derived VCs.
FreeGSNKE VCs are calculated using finite differences of GS solutions.

We explore and compare several setups to calculate the VCs from the emulators. 
First, we use automatic differentiation (AD) to obtain the shape matrix in eq.~\ref{eq:linearisation_jacobian}.
For this, we compare using the single best model, as measured by cross-validation MSE loss, to using the ensemble of eight models. 
Second, we compare the effect of using VCs derived from automatic differentiation with those constructed by finite difference (FD) differentiation, both using the ensemble of eight models. 
Finally, we compare the same to the numerically calculated FreeGSNKE VCs.

A compact quantitative summary of these comparisons is given in \cref{tab:VC_summary}.
The table reports representative diagonal errors, diagonal spreads, and off-diagonal couplings for each VC construction method, with the full realised shift matrices provided in \cref{app:all_tables}.
The NN ensemble-based VCs achieve a typical diagonal errors and diagonal standard deviation values comparable to the numerically calculated FreeGSNKE baseline, with similar typical off-diagonal coupling of $0.42\%$.
The largest off-diagonal couplings occur for quantities related to the divertor, especially realised shifts in $R_\text{strike}$, consistent with the more detailed discussion below.

\begin{table}[b!]
\centering
\footnotesize
\setlength{\tabcolsep}{3pt}
\renewcommand{\arraystretch}{1.15}
\caption{
Summary of realised shift statistics for the main VC construction methods.
All numerical entries are expressed as percentages of the requested \SI{5}{\milli\metre} shift.
Here, ``diag.\ error'' denotes $\mathrm{median}_i|\mu_{ii}-1|\times 100$, where $\mu_{ii}$ is the mean normalised realised shift when parameter $i$ is requested,
``diag.\ std.'' denotes $\mathrm{median}_i\,\sigma_{ii}\times 100$, where $\sigma_{ii}$ is the standard deviation of the normalised realised shift for requested parameter $i$,
``off-diag.\ coupling'' denotes $\mathrm{median}_{i\neq j}|\mu_{ij}|\times 100$, quantifying the typical unwanted cross-talk between shape parameters,
and ``max.\ off-diag.\ coupling'' denotes $\max_{i\neq j}|\mu_{ij}|\times 100$.
A perfect VC would yield diag.\ error and diag.\ std.\ of zero, and vanishing off-diagonal coupling.
}
\label{tab:VC_summary}
\begin{tabularx}{\linewidth}{@{}
>{\raggedright\arraybackslash}p{0.16\linewidth}
>{\centering\arraybackslash}p{0.12\linewidth}
>{\centering\arraybackslash}p{0.12\linewidth}
>{\centering\arraybackslash}p{0.13\linewidth}
>{\centering\arraybackslash}p{0.13\linewidth}
>{\raggedright\arraybackslash}X
@{}}
\toprule
VC construction
&
Diag. error [\%]
&
Diag. std. [\%]
&
Off-diag. coupling [\%]
&
Max. off-diag. coupling [\%]
&
Notes \\
\midrule

Single NN, AD
& 0.7
& 4.7
& 0.45
& 14.7
& Lowest latency emulator case; broader distributions than the ensemble. \\

NN ensemble, AD
& 1.6
& 3.1
& 0.42
& 10.9
& Reduced variance relative to the single model. \\

NN ensemble, FD
& 1.3
& 3.2
& 0.43
& 20.7
& Uses medium current shifts from \cref{tab:fin_diff_shift_floors}; comparable to AD. \\

FreeGSNKE FD GS
& 2.7
& 2.1
& 0.51
& 24.7
& Physics-based reference; narrowest diagonal spread, with residual divertor coupling. \\

\bottomrule
\end{tabularx}
\end{table}

In \cref{fig:small_grid_histograms_both}, we show histograms of the realised shifts for a subset of the controlled shape parameters, focussing on $R_\text{out}$ and $R_\text{strike}$.
These histograms illustrate the same behaviour summarised quantitatively in \cref{tab:VC_summary}: the core shape parameters are accurately decoupled, while the largest residual couplings are associated with divertor leg and strike point control.
In the upper-left panel, for example, the shift in $R_\text{out}$ is displayed after shifts to the coil currents have been applied corresponding to a requested shift of the same shape parameter.
In the upper right, the realised shift in $R_\text{strike}$ is shown for the same requested shift of 5 mm in $R_\text{out}$.
All histograms are normalised to the requested shift size of 5 mm. 
Corresponding histograms for all 7 shape parameters are provided in \cref{app:all_histograms}.
Those displayed in \cref{fig:small_grid_histograms_both} illustrate the best- and worst-performing VCs.
In all of Figs~\ref{fig:small_grid_histograms_both}, \ref{fig:pred_shifts_hists_all_params_finte_vs_auto} and \ref{fig:pred_shifts_hists_all_params_FGS_ensemble_single}, the vertical axes in panels on the diagonal are centred around $1,$ corresponding to a perfect match of the requested and realised shifts.
The off-diagonal panels have vertical axes centred around $0$, corresponding to the desired orthogonality of the realised shifts. 

\subsection{Single model vs ensemble prediction}

\begin{figure}[h!t]
    \centering
    \begin{subfigure}{0.49\linewidth}
        \includegraphics[width=0.99\textwidth]{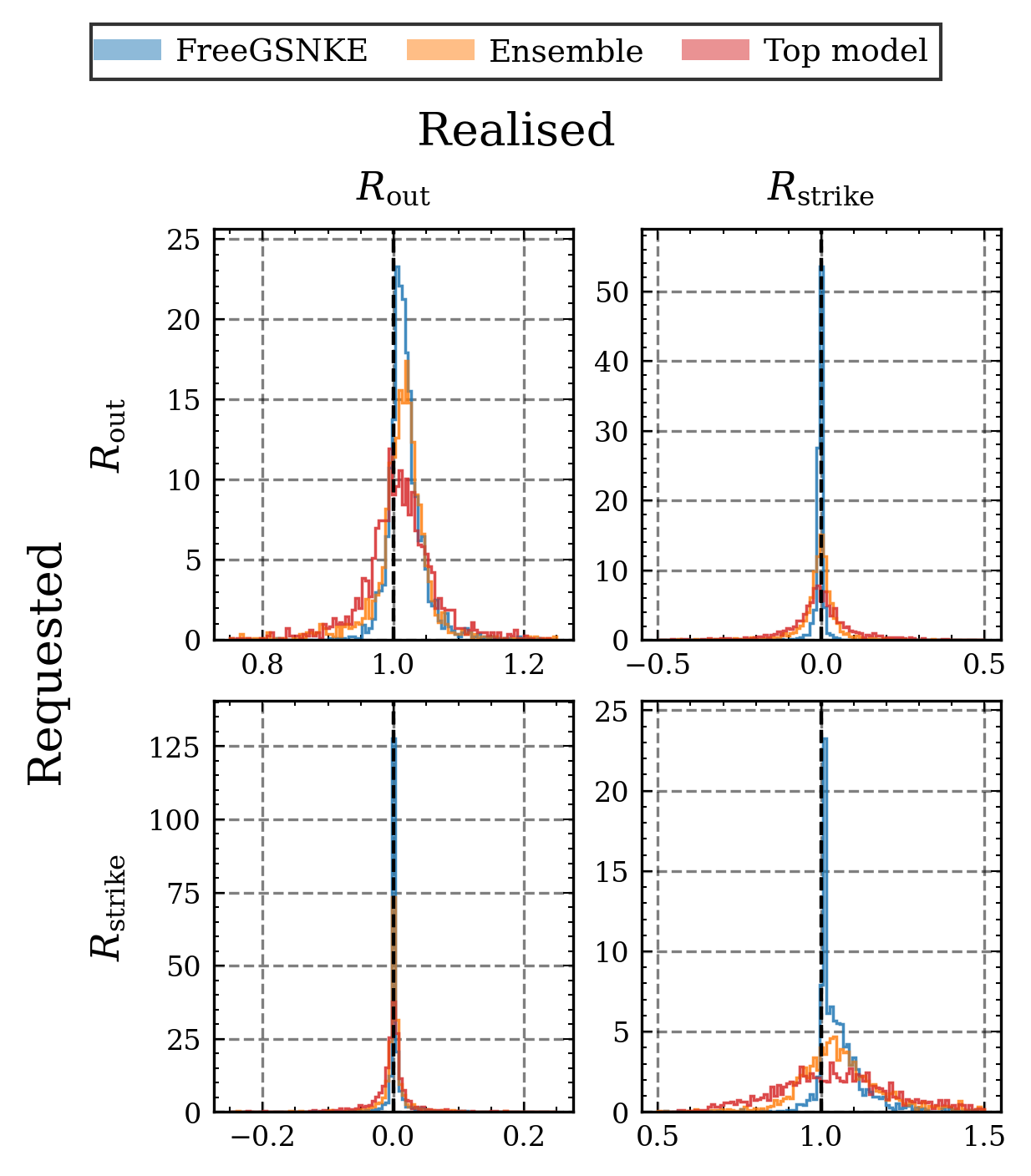}
        \caption{}
        \label{fig:pred_shifts_hists_smallgrid_fgs_ensemble_single}
    \end{subfigure}
    \begin{subfigure}{0.49\linewidth}
        \includegraphics[width=0.99\textwidth]{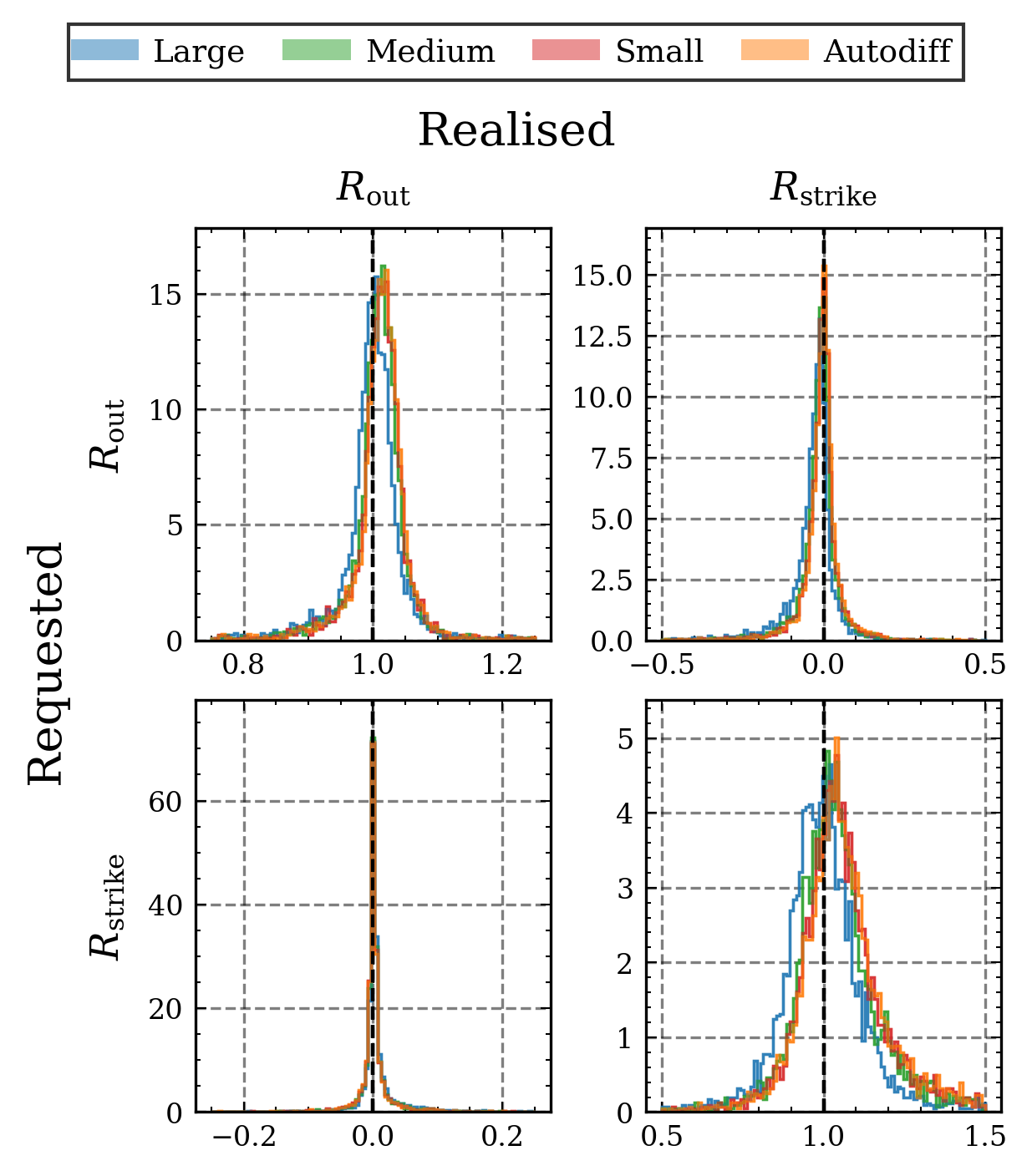}
        \caption{}
    \label{fig:pred_shifts_hists_smallgrid_finite_vs_auto}
    \end{subfigure}
    \caption{
    Histograms of realised shifts, as in \cref{eq:requested_shift}, for $R_\text{out}$ and $R_\text{strike}$, normalised to the requested shift $\delta \bm P _\text{req}$. 
    \textit{Left panels} (a): finite-difference GS derivatives (blue) with emulator VCs from the ensemble (orange) and the top model (red), evaluated on early-chain equilibria (MCMC step $< 50$) within the test set.
    This shows that the FreeGSNKE VCs perform best, and that an ensemble of models performs better than the single top model. 
    \textit{Right panels} (b): comparison of emulator automatic-differentiation-based VCs (orange) vs finite-difference VCs (blue, green, red) for the ensemble of models, for different coil current shifts as per \cref{tab:fin_diff_shift_floors}.
    Further histograms of this form, for all 7 emulated shape parameters, are shown in \cref{fig:pred_shifts_hists_all_params_FGS_ensemble_single} in \cref{app:all_histograms}.}
    \label{fig:small_grid_histograms_both}
\end{figure}

The histograms in \cref{fig:pred_shifts_hists_all_params_FGS_ensemble_single} in \cref{app:all_histograms} show that the ensemble of the top-eight models is consistently more accurate than the single best-performing model across all the shape parameters, as shown by the narrower, more peaked distributions.
This is also visible in the diagonal plots in \cref{fig:pred_shifts_hists_smallgrid_fgs_ensemble_single} where the distribution from the top model is wider than the ensemble, although good performance is still demonstrated by both approaches. 
The means and standard deviations of the top model for all shape parameters are reported in \cref{tab:pred_means_stdev_top_model}, as well as those for the ensemble in \cref{tab:pred_means_stdev_ensemble_auto} in \cref{app:all_tables}. We use half the difference between the 16\textsuperscript{th} and 84\textsuperscript{th} percentile as an estimate of the standard deviation to be more robust to outliers. 
The top model has errors in the range of a few percent, mostly between $2\%$ and $9\%$ for most shape parameters, while the ensemble errors are typically below $5\%$. 
The realised shifts in $R_\text{strike}$ have larger errors for both the top model and the ensemble, up to $25\%$ and $15\%$ respectively, as well as a slightly displaced peak visible in the bottom right plot in the figure.

\subsection{Emulator finite differences vs automatic differentiation}
\label{sec:finite_diff_vs_auto}

We compare the performance of VCs obtained from the emulators using both automatic differentiation and finite differences.
Automatic differentiation provides a natural and efficient route, as derivatives of the trained NN functions are readily available in modern machine learning frameworks.
However, the models were not explicitly trained to reproduce shape sensitivities, and their local smoothness is not guaranteed a priori.
We investigate these effects on the reliability of derivatives obtained via automatic differentiation.

Finite-difference-based derivatives provide a complementary estimate to automatic-differentiation-based derivatives, in which the perturbation size used to compute the finite differences can be tuned to the accuracy of the emulator in the input parameters.
This enables a more controlled evaluation of the local response, ensuring that the inferred derivatives reflect physically meaningful variations rather than small-scale emulator noise.
This comparison allows us to assess whether the emulator-derived sensitivities are sufficiently accurate and stable for use in control, and whether automatic differentiation can be reliably used in place of finite-difference estimates.

For the finite-difference calculation, we explore the use of a set of different current shifts, as in  \cref{tab:fin_diff_shift_floors} in \cref{app:coil_current_shift_values}.
The `medium' size is calibrated so that each coil current change results in a plasma response of approximately two MAE in at least one of the considered shape parameters, for a representative MAST-U equilibrium.
This is to ensure that the finite-difference-based derivatives are calculated on physically meaningful emulated plasma responses, rather than on small scale emulator noise.
We use these exact current shifts to calculate the finite-difference-based derivatives of the emulators in \cref{eq:linearisation_jacobian} for all \num{5000} equilibria considered in this validation. 

\Cref{tab:pred_means_stdev_ensemble_auto} shows the means and standard deviations of the normalised realised shifts for the automatic-differentiation-based VCs.
The performance is comparable for both approaches for the core shape parameters, with errors in the range of $1\% - 5 \%$.
This shows that the use of a single equilibrium to calibrate the shifts used in finite differences does not appear to have a detrimental effect on performance, even when applied to very different equilibria.
At the same time, the close agreement between the automatic differentiation and finite difference results in \cref{tab:VC_summary} indicates that the emulators are sufficiently smooth for either approach to be used in the calculation of the shape matrix.

The accuracy relating to the realised shifts in the strike point radius is less optimal, with a standard deviation of approximately
 $12\%$ for the requested $R_\text{strike}$ shift, approximately $8\%$ for the requested $R_\text{X}$ shift, and approximately
 $15\%$ for the realised $R_\text{strike}$ when requesting shifts in $R_\text{nose}$.
 These figures are consistent across both automatic-differentiation-based and finite-difference-based derivatives.
 This pattern persists when measuring FreeGSNKE numerical results (as discussed in the next subsection), showing that this behaviour is not unique to emulator-based VCs.
This decreased orthogonality and increased variance when aiming to isolate control of $R_\text{strike}$, $R_\text{X}$ and $R_\text{nose}$ reflects the intrinsic difficulty of decoupling these specific divertor shape parameters with the available MAST-U coil set.
In particular, these radial descriptors exhibit strong mutual coupling and are therefore inherently more challenging to control independently.
This is further supported by \cref{fig:pearson_output}, which demonstrates strong correlation between these parameters and is consistent with operational practice, where such parameters are rarely all regulated simultaneously in feedback. 

\subsection{Grad Shafranov finite differences vs neural network Jacobians}
\label{sec:Freegsnke_vs_emulators}
Having established the relative performance of the different approaches for computing emulator-based VCs, we now benchmark them against VCs obtained from finite-difference GS solutions in FreeGSNKE.
We refer to these as GS VCs, and the corresponding realised shifts as GS-realised shifts.
\Cref{fig:pred_shifts_hists_smallgrid_fgs_ensemble_single} shows the GS-realised shifts alongside those obtained from the NN-based VCs for $R_\text{out}$ and $R_\text{strike}$, with corresponding full summary statistics for all shape parameters reported in \cref{tab:pred_means_stdev_FreeGSNKE}.
The compact comparison in \cref{tab:VC_summary} shows that the GS VCs have the smallest typical diagonal standard deviation.
As expected, the GS VCs therefore provide the most precise realisation of the requested shifts.
However, the emulator-based VCs show strong agreement across the majority of shape parameters, particularly for core quantities, with only modest degradation in accuracy.
This modest loss in accuracy is offset by a substantial reduction in computational cost: while GS-based VCs require repeated equilibrium solves, the emulator-based VCs can be evaluated almost instantaneously.
This enables state-aware VC evaluation at timescales compatible with real-time control, which is not feasible with conventional approaches.

The differences are most pronounced for divertor-related parameters, especially the strike point, where larger deviations are observed.
However, this behaviour is also present in the GS-based VCs, reflecting the intrinsic difficulty of decoupling these quantities with the available MAST-U coil set.

Overall, the emulator-derived VCs provide a close approximation to the physics-based reference, capturing the essential structure of the shape sensitivities while enabling real-time applicability through dramatically reduced latency.

\begin{figure}[h!t]
    \centering
    \includegraphics[width=0.9\linewidth]{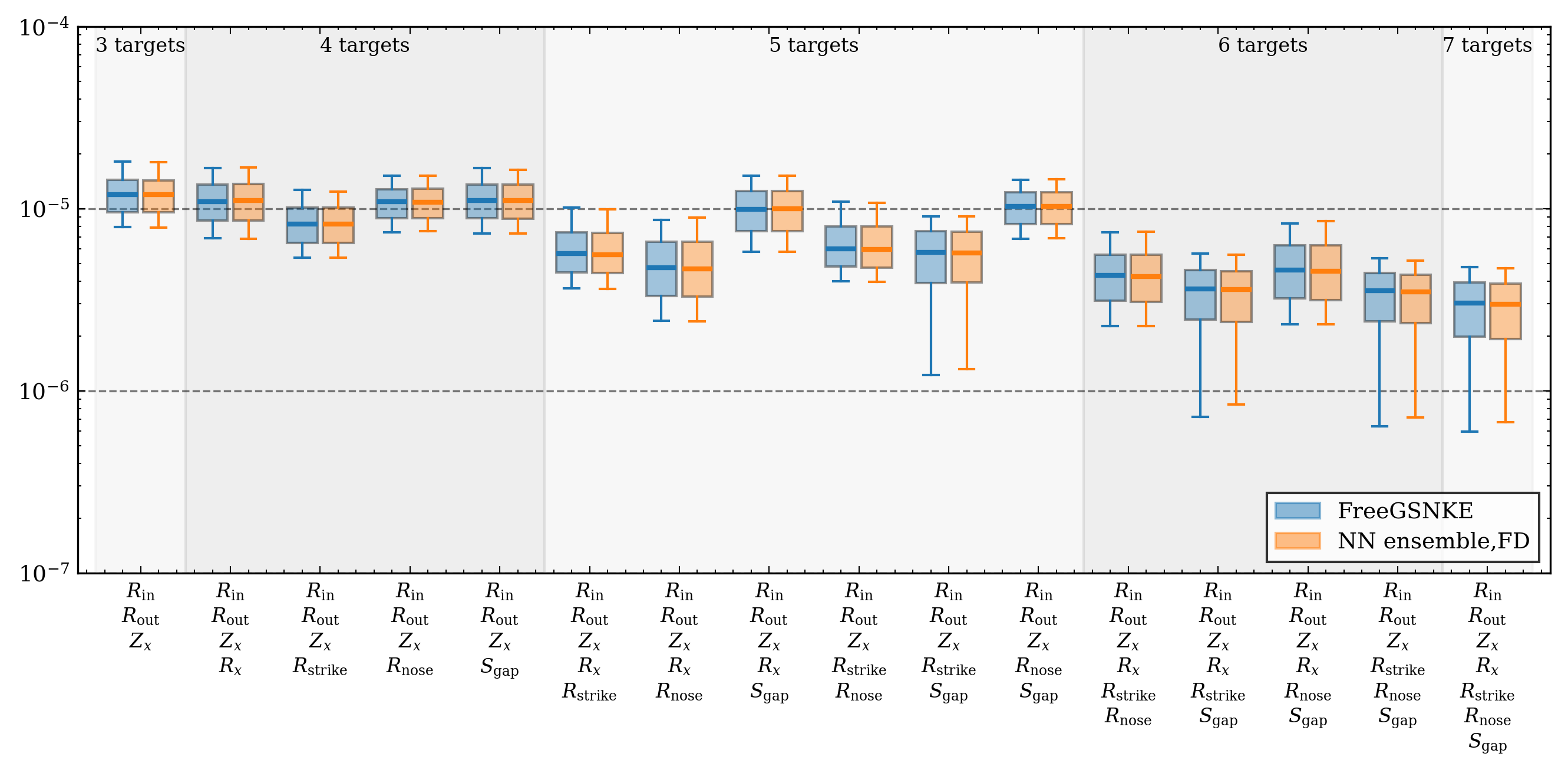}
    \caption{Box plots showing the smallest singular values in the decomposition of the sensitivity matrix for finite difference GS derivatives and finite differences of the ensemble of models, evaluated on early-chain equilibria (MCMC step $<50$) within the test set.
    For the subsets shown, the smallest singular values decrease as more shape parameters are included, with similar distributions for the GS and emulator matrices.}
    \label{fig:singular_values}
\end{figure}

Moore--Penrose inversion is applied to both the GS and emulator sensitivity matrices, thresholding singular values at $\sigma_k > 10^{-15}\sigma_{\max}$, where $\sigma_{\max}$ is the largest singular value, with no additional truncation to select fewer control directions.
\Cref{fig:singular_values} compares the smallest singular values for different subsets of shape parameters on early-chain test equilibria (MCMC step $<50$).
The smallest singular values decrease as more shape parameters are included, with similar distributions for the GS and emulator matrices.
Together with the realised shift results in \cref{tab:VC_summary}, this agreement supports the reproduction of the GS sensitivity structure and approximate decoupling in the demonstrated regime.
Weakly responsive directions can nevertheless require large current increments because their SVD gains scale as $1/\sigma_k$.
More restrictive truncation, as used in SVD-based output regulation to trade tracking accuracy against control effort \cite{ambrosino2007steady,ambrosino2008xsc}, is compatible with either source of sensitivity matrices.

The sampled coil currents predominantly lie within operating limits, as the MCMC algorithm penalises excursions beyond them (\cref{sec:dataset_generation,fig:input_feature_distributions}).
This sampling preference does not ensure that every VC request respects the remaining current headroom.
Current and voltage limits are handled downstream in the MAST-U PCS \cite{MCARDLE2020111764} and in the virtual PCS  used for the companion closed-loop evaluation \cite{dynamic_val_paper}.
Coil current saturation was not observed to limit shape control in those evaluated scenarios.

\section{Discussion and conclusion}
\label{sec:discussion}

In this work, we have presented an NN framework for deriving VCs for use in real-time plasma shape control using differentiable emulators of plasma shape parameters.
We have shown that the resulting VCs reproduce the behaviour of finite-difference GS linearisation with high fidelity across a wide range of plasma configurations, establishing this approach as a route to replacing offline, precomputed VC schedules with state-aware VCs evaluated directly during plasma control.
Work is ongoing to integrate and deploy this capability within a real-time plasma control environment.

A key practical advantage of the proposed approach is the speed of NN inference relative to conventional finite difference GS linearisation.
Because NNs can evaluate shape parameters and their Jacobians orders of magnitude faster than a conventional GS solver, it opens the possibility of providing controllers with state-aware VCs in real time.
This represents a fundamental shift from the current paradigm of precomputed, piecewise-constant VC schedules, and allows the control system to remain effective even as the plasma evolves away from nominal operating conditions.

The accuracy of the emulator-derived VCs is excellent for core shape parameters, with post-VC shift errors typically in the range of $1\%-5\%$.
Larger deviations are observed for divertor-related quantities, particularly the strike point, where errors can reach $\sim 15\%$.
This reflects the intrinsic difficulty of decoupling these parameters with the available coil set, rather than a limitation of the approach itself.
We also observe that the use of an ensemble of models reduces variance and improves robustness compared to single-network predictions.

The emulator supplies an equilibrium-dependent sensitivity matrix, while the choice of usable control directions remains part of controller design.
The seven shape descriptors and ten shaping coil current inputs used here give a $7 \times 10$ sensitivity matrix, but this nominal actuator count does not guarantee seven practically usable directions.
Weakly responsive directions require large current increments because their SVD gains scale as $1/\sigma_k$.
For the static comparison, both GS and emulator matrices are pseudoinverted with the same numerical cutoff of $\sigma_k > 10^{-15}\sigma_{\max}$ without additional truncation.
A more restrictive truncation can be applied to either matrix to reduce control effort, at the expense of tracking accuracy in the discarded directions.
The emulators therefore inherit the conventional trade-off between tracking accuracy and control effort.

The nonlinear GS tests support approximate decoupling for the selected equilibria and requested shifts.
They do not establish that arbitrary shape requests are feasible within coil limits.
For example, the minimum-norm pseudoinverse does not constrain individual coil currents.
Current and voltage limits are enforced downstream in the MAST-U PCS and in the virtual PCS of the companion closed-loop study \cite{dynamic_val_paper}.
Coil current saturation was not observed to limit shape control in those evaluated scenarios.

Further improvements to the presented approach to shape matrix emulation are possible.
The accuracy of strike point control could be enhanced by refining the training dataset to include a larger number of highly diverted equilibria, thereby reducing prediction errors in challenging regions of parameter space.
In addition, the computation of derivatives -- and hence virtual circuits -- could be improved through the use of Sobolev training \cite{DBLP:journals/corr/CzarneckiOJSP17}, in which derivative information is included in the loss function used for NN optimisation.
Shape parameter gradient information could also be added for a small subset of training data on which NNs pre-trained for regression are fine-tuned.
Such an approach may enable improved sensitivity estimation without the need to train directly on large derivative datasets, offering a practical route to further performance gains.

In addition to enabling real-time operation, the proposed approach simplifies controller deployment by reducing the need for manual design of VC schedules.
In a prospective real-time implementation, the linearisation point would be specified by the supplied active coil currents, total plasma current and profile parameters $(\bm I_\mathrm{act}, I_p, \bm\theta)$.
The sensitivity matrix would be evaluated at this point by automatic differentiation or finite differences with respect to the shaping currents, holding the other inputs fixed, and pseudoinverted to obtain the VCs.
The resulting matrix would map the shape adjustments requested by the controller to coil current adjustments within the existing PCS, as described in \cref{sec:VC_intro}.
Deployment therefore requires external provision of the equilibrium inputs.
In particular, when real-time estimates of the four profile coefficients are unavailable, they can be prescribed from a reference discharge with comparable heating and fuelling~\cite{dynamic_val_paper}.
Coupling state estimation to VC evaluation is a possible extension, whereas the present study validates the VCs for supplied equilibrium inputs and does not assess the effect of errors in an inferred state.
The method described in this work alleviates the requirement for expert selection of reference equilibria and tuning of control phases, while retaining the familiar and interpretable VC-based control structure.

The generality of the methodology concerns learning a forward mapping from equilibrium inputs to selected shape descriptors, differentiating that mapping and constructing VCs from the resulting sensitivity matrix.
The descriptor set and control architecture are application-specific design choices.
Increasing the number of descriptors does not itself add actuator degrees of freedom or guarantee better decoupling.
Applying the procedure to another device or extending the model to additional descriptors requires corresponding equilibrium data, model training and validation.
The trained models presented here remain specific to the MAST-U inputs and seven outputs used in this study, but the methodology is inherently device-agnostic.

The emulator-derived shape matrices can also provide local shape sensitivities to control frameworks other than the VC formulation tested in this work.
For example, the same matrices could be used in XSC-like output regulation, with output/current weighting and singular-value truncation applied downstream~\cite{ambrosino2008xsc}.
Such use is limited to the descriptors predicted by the emulators and requires controller design and validation for the chosen framework.

Overall, this work demonstrates that neural-network-based VC computation can retain the structure and interpretability of conventional control methods while enabling real-time, state-aware operation.
This combination of accuracy, speed, and reduced reliance on manual tuning makes it a promising route toward more adaptive and accessible plasma control in future fusion devices.

\clearpage

\funding{This work was funded by the Fusion Computing Lab collaboration (between UKAEA and STFC Hartree Centre) and was part funded by the EPSRC Energy Programme (EP/W006839/1).}

\data{Data used for neural network optimisation and testing is available at \url{https://doi.org/10.14468/rrbr-vf87.}}

\clearpage

\appendix

\clearpage
\section{Seeding shot descriptions}
\label{app:seeding_shot_descriptions}
In this appendix section, we list the shots used for the MCMC seeding equilibria, along with descriptions of each scenario. 
Shot descriptors adopt the structure \textit{core configuration -- plasma current (kA) -- divertor configuration -- NBI heating}.
Core configurations can be: \texttt{DN} (double null), \texttt{LIM} 
(limited), \texttt{LSN} (lower single null), or \texttt{USN} (upper single null). 
Divertor configurations can be: \texttt{CD} (conventional divertor), \texttt{SXD} (Super-X divertor), \texttt{LIM} (limited), \texttt{SF} (snowflake), or \texttt{XD} (X-divertor). 
NBI heating can be: \texttt{OH} (purely Ohmic), \texttt{1BSS} (south beam), \texttt{1BSW} (south-west beam), or \texttt{2B} (both beams).

\begin{table}[h!]
\centering
\caption{Shots used to generate seeding equilibria for MCMC dataset generation.}
\label{tab:seed_shot_descriptions}
\begin{tabular}{|l|l|}
\hline
Shot(s) & Scenario \\
\hline\hline
45272, 47642, 49107, 49321, 49347, 49404 & DN-750-CD-2B \\
48194, 48912                             & DN-750-CD-1BSS \\
48286                                    & DN-750-CD-1BSW \\
48775, 49308, 49382                      & DN-750-CD-OH \\
49062, 49274, 49326                      & DN-750-SXD-2B \\
47417                                    & DN-750-SXD-OH \\
49449                                    & DN-750-SXD-1BSW \\
49416                                    & DN-1000-CD-2B \\
48303                                    & DN-600-CD-1BSS \\
48304                                    & DN-600-CD-1BSW \\
48701                                    & DN-600-CD-OH \\
49463                                    & DN-600-CD-2B \\
45352                                    & DN-600-SXD-1BSW \\
45371                                    & DN-600-SXD-OH \\
49374                                    & DN-600-XD-1BSW \\
49467                                    & DN-600-SF-2B \\
48039, 48680                             & DN-450-CD-OH \\
49170                                    & DN-450-CD-2B \\
48038                                    & DN-450-SXD-OH \\
49213                                    & LSN-750-CD-2B \\
49262                                    & USN-750-SXD-2B \\
46939                                    & LIM-750-LIM-OH \\
\hline
\end{tabular}
\end{table}

\clearpage
\section{Synthetic data distributions}
\label{sec:synthetic_data_distributions}
This appendix section provides supplementary diagnostics for the synthetic equilibrium dataset introduced in \cref{sec:dataset_generation} and used to train the NN shape emulators.
The purpose of these plots is to make explicit the domain over which the networks are asked to interpolate, and to show how the training set covers the combinations of coil currents, plasma current, profile parameters, and shape parameters that define the MAST-U equilibrium library.
All quantities in this appendix are shown in their physical or native unstandardised coordinates, before the input and output standardisation described in \cref{sec:dataset_generations} is applied for NN training.

\begin{figure}[h]
    \centering
    \includegraphics[width=1\linewidth]{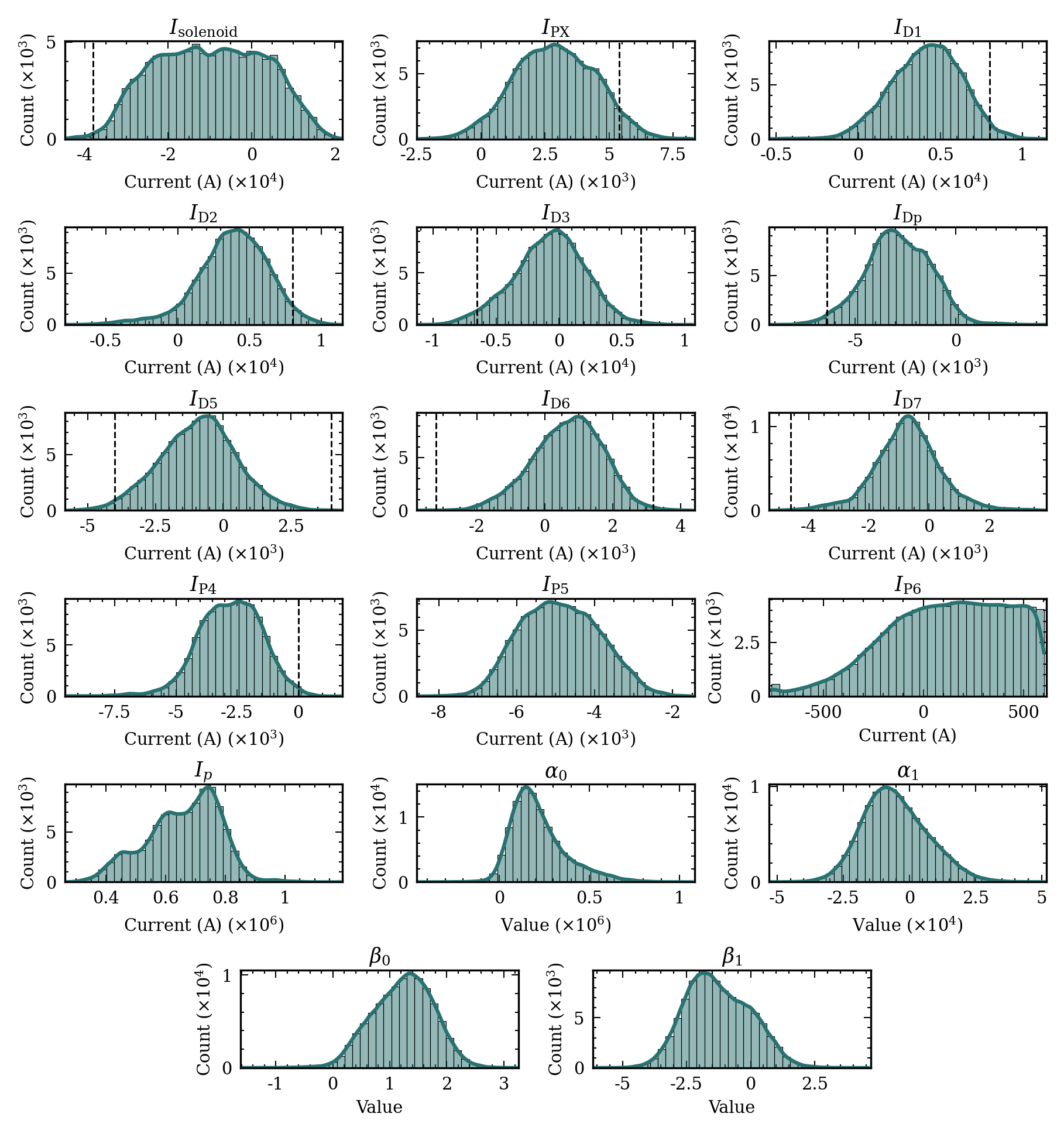}
    \caption{
    Distributions of the input features used to train the NN emulators.
    Black dashed vertical lines indicate the current limits for the solenoid and shaping coils where they fall within the plotted range.
    }
    \label{fig:input_feature_distributions}
\end{figure}

\Cref{fig:input_feature_distributions} shows the marginal distributions of the model inputs.
The PF coil current distributions are generally broad and single-peaked, reflecting the MCMC exploration around experimentally relevant MAST-U operating points while retaining a strong preference for current configurations away from machine limits.
Dashed vertical lines indicate the shaping coil and solenoid current limits within the plotted ranges.
The plasma current distribution is similarly concentrated around the range of values represented by the seeding shots and accepted synthetic equilibria.
These input histograms are also useful for identifying future real-time queries that may lie outside the learned domain.

\begin{figure}[h]
    \centering
    \includegraphics[width=1\linewidth]{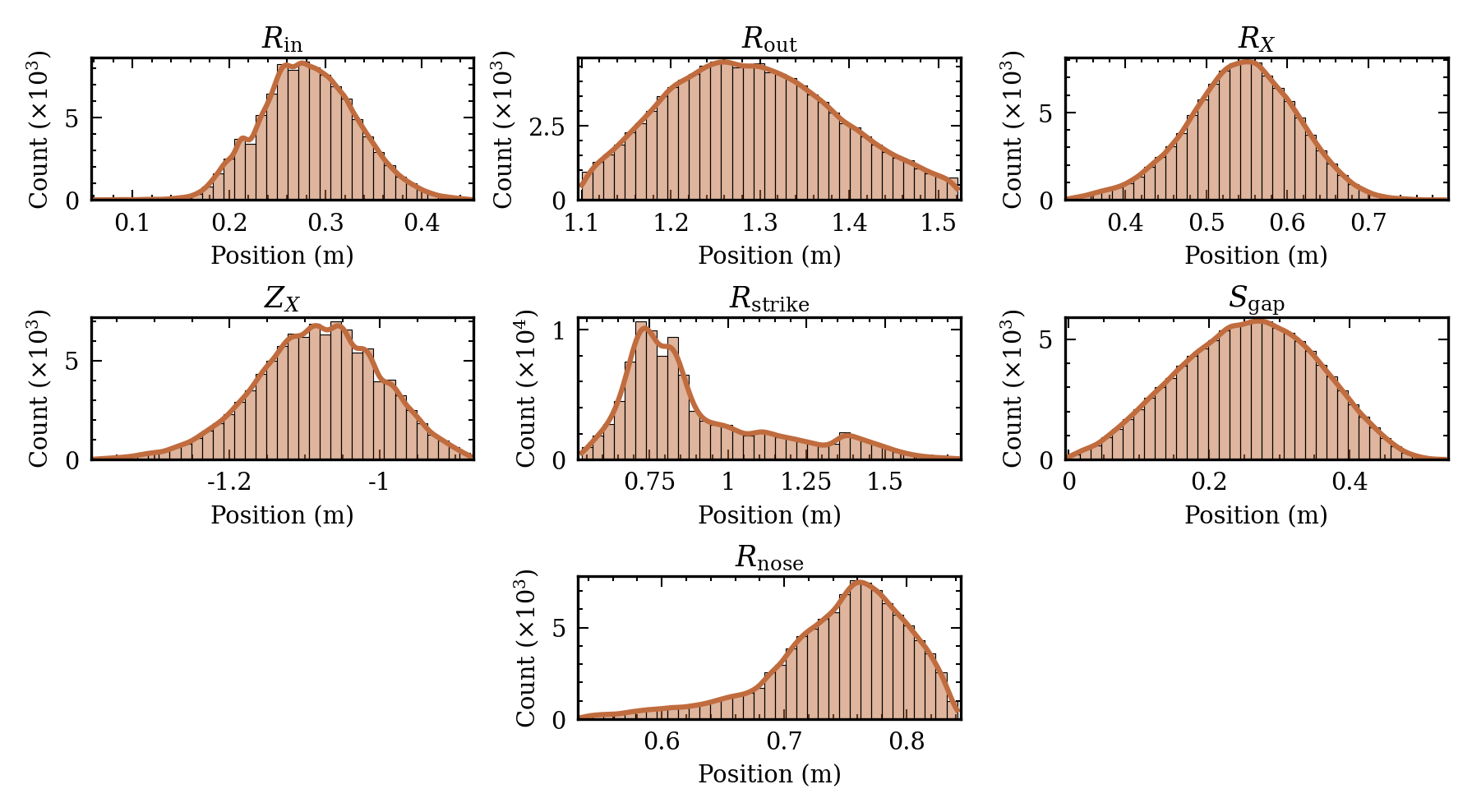}
    \caption{
    Distributions of the output shape parameters used to train the NN emulators.
    These distributions correspond to the accepted synthetic equilibria after the strike point consistency filtering described in \cref{sec:dataset_generation}.
    }
    \label{fig:output_feature_distributions}
\end{figure}

The corresponding output distributions are shown in \cref{fig:output_feature_distributions}.
These histograms demonstrate that the synthetic library spans a wide range of core and divertor shape parameters.
Several output distributions are non-Gaussian or asymmetric, particularly those associated with divertor geometry.
The breadth and non-normalness of these output distributions are important for the present application: the emulator is not only required to reproduce a narrow family of equilibria, but also to provide smooth shape derivatives across a substantial range of plausible MAST-U shapes.

\begin{figure}[h]
    \centering
    \includegraphics[width=1\linewidth]{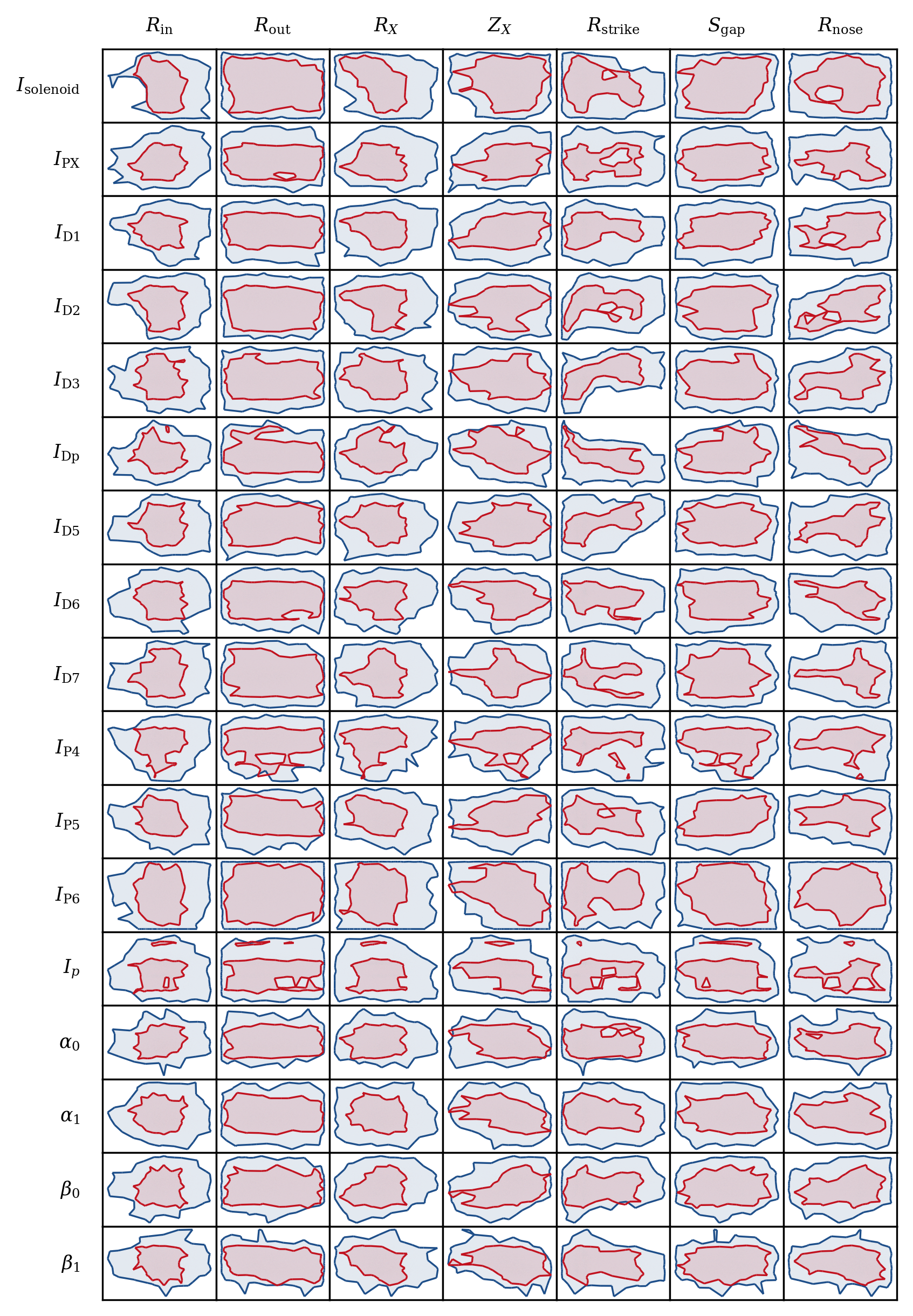}
    \caption{
    Two-dimensional coverage of the training dataset for every model input and output pair.
    Rows correspond to model inputs and columns correspond to the output shape parameters.
    In each panel, the shaded alpha-shape region gives an approximate boundary of the sampled support in that two-dimensional projection, with the red region indicating the seed points and the blue region indicating the accepted MCMC samples.
    The area covered by the MCMC samples is generally larger than that covered by the seed points, indicating that the MCMC procedure has successfully expanded the training dataset beyond the initial seeding equilibria.
    }
    \label{fig:input_output_alpha_grid}
\end{figure}

The marginal histograms alone do not show whether all combinations of inputs and outputs are represented.
For this reason, \cref{fig:input_output_alpha_grid} shows two-dimensional input-output coverage for every pair of model input and output variables.
The shaded alpha-shape regions indicate the approximate support of the training data in each two-dimensional projection.
These plots reveal the parts of input-output space in which the network is constrained by data, as opposed to regions where a queried equilibrium would require extrapolation.
The shapes of the regions also show that the library is structured rather than rectangular: physically feasible equilibria occupy correlated subregions of the input and output ranges.

\begin{figure}[h]
    \centering
    \includegraphics[width=1\linewidth]{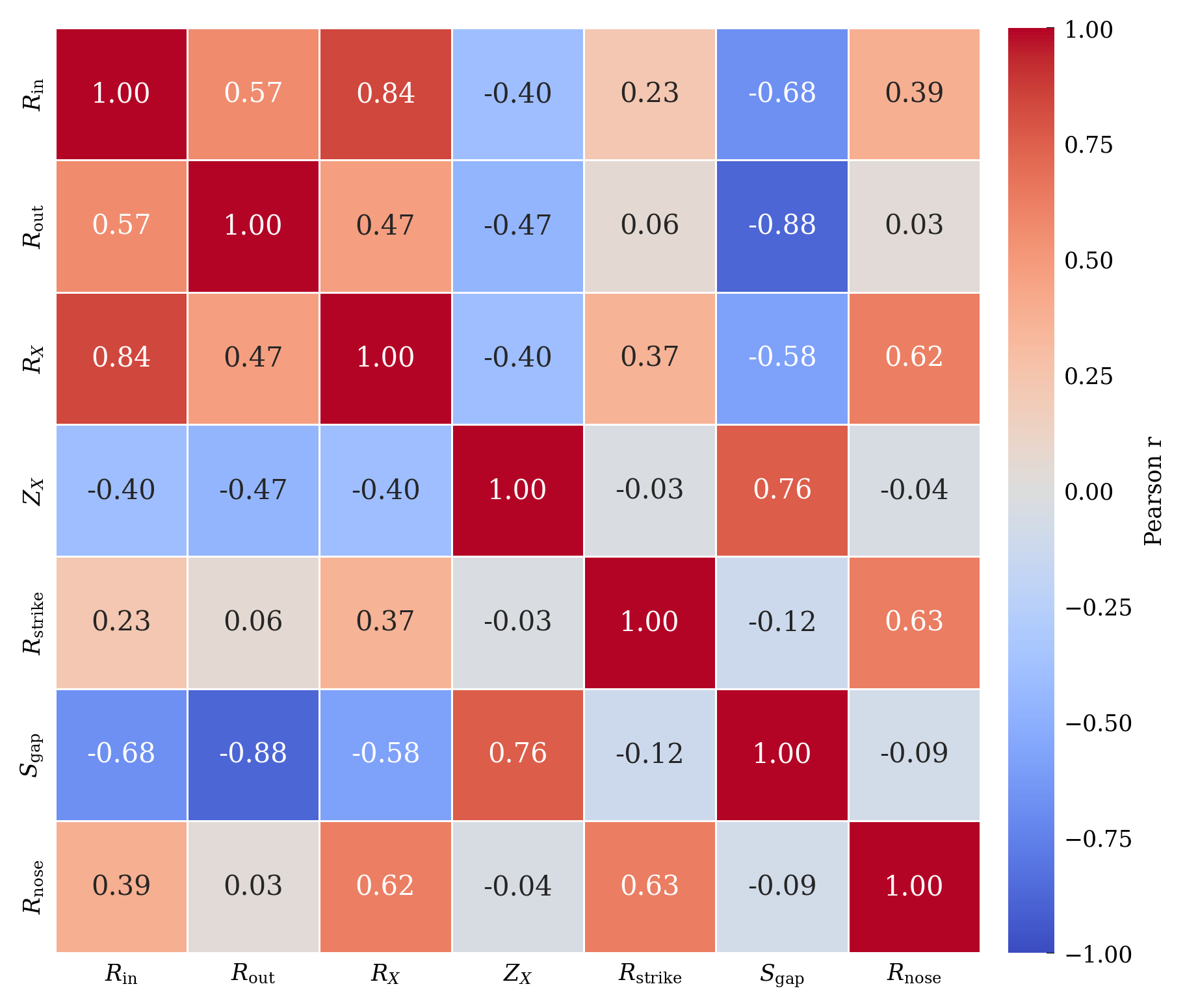}
    \caption{
    Pairwise Pearson correlation coefficients among the output shape parameters in the training dataset.
    The colour scale gives the Pearson coefficient $r$.
    The strongest correlations highlight geometric coupling within the sampled equilibrium library, including anticorrelations between $S_\text{gap}$ and the core radial quantities, a positive association between $S_\text{gap}$ and $Z_\text{X}$, and correlations involving $R_\text{nose}$, $R_\text{strike}$, and $R_\text{X}$.
    These global correlations motivate the need for VC-based decoupling, while also providing a summary of the output-space structure learned by the emulator.
    }
    \label{fig:pearson_output}
\end{figure}

Finally, \cref{fig:pearson_output} shows the pairwise Pearson correlations among the seven output shape parameters.
The correlations should not be interpreted as a complete sensitivity matrix, since they are global, linear, and computed from the sampled data distribution.
Nevertheless, the output correlations emphasise that the shape control problem is intrinsically coupled.
In particular, $S_\text{gap}$ is strongly anticorrelated with several core radial quantities, especially $R_\text{in}$ and $R_\text{out}$, and is positively correlated with $Z_\text{X}$.
The radial X-point and midplane quantities also show appreciable correlations with one another, while $R_\text{nose}$ is positively correlated with $R_\text{strike}$ and with some core radial parameters.
These correlations are consistent with the geometric interpretation of the outputs: changes in the overall plasma size, radial position, X-point location, and divertor-leg geometry cannot generally be varied independently in the sampled equilibrium set.
They also motivate the use of VCs, since a controller acting directly on PF coils must compensate for these coupled responses in order to produce approximately orthogonal shape-parameter shifts.

\clearpage
\section{Trained models architecture and hyperparameters}
\label{app:model_architectures}

\Cref{tab:top_models} lists the model and training hyperparameters used in the ensemble of predictors.
The rank column refers to the models performance when ranked by cross-validation MSE during hyperparameter search.

\begin{table}[htbp] 
\centering 
\caption{Hyperparameters of the Top-Ranked Models} \label{tab:top_models} 
\begin{tabular}{c c c l l l c} 
\hline Rank & Width & Depth & Activation & Optimiser & Kernel Initializer & Batch Size \\ 
\hline 
1 & 256 & 3 & sigmoid & Adam & HeNormal & 64 \\ 
2 & 256 & 4 & celu & Adam & RandomNormal & 32 \\ 
3 & 256 & 4 & sigmoid & Adam & HeNormal & 256 \\ 
4 & 256 & 3 & sigmoid & Adam & GlorotNormal & 128 \\ 
5 & 256 & 4 & sigmoid & Adam & HeNormal & 256 \\ 
6 & 256 & 3 & sigmoid & Adam & GlorotNormal & 256 \\ 
7 & 128 & 5 & sigmoid & Adam & RandomNormal & 128 \\ 
8 & 128 & 3 & sigmoid & Adam & RandomNormal & 64 \\ 
\hline 
\end{tabular} 
\end{table}

\clearpage
\section{Model prediction performance}
\label{app:model_prediction_performance}
\begin{figure}[h!t]
    \centering
    \includegraphics[height=0.85\textheight]{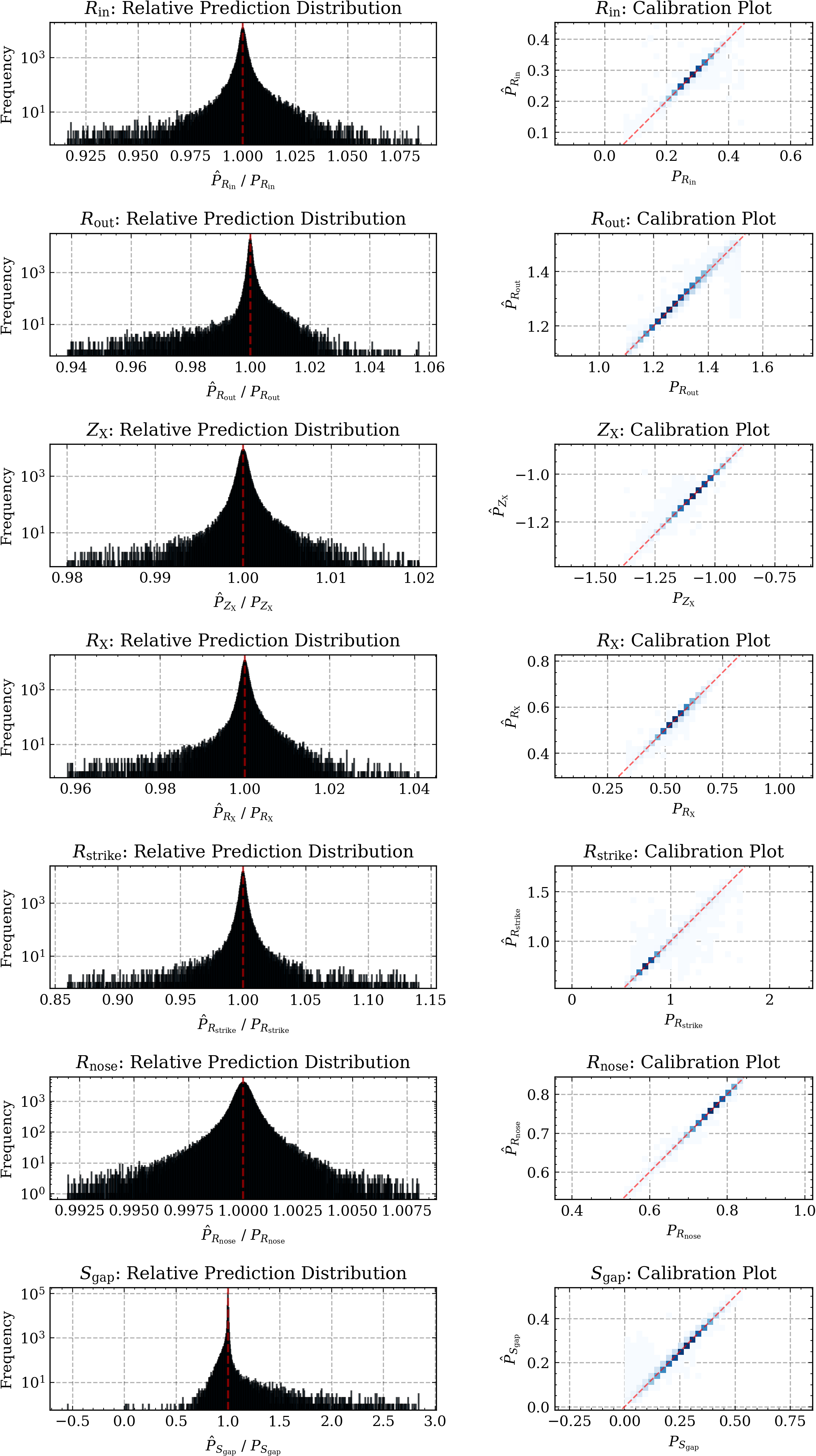}
    \caption{Ensemble model performance evaluated on the test set. On the left are histograms of the model predictions relative to the true value for each shape parameter output. On the right are 2-dimensional histograms of predicted value versus true value for each shape parameter.
    }
    \label{fig:models_forward_pred}
\end{figure}

In this appendix section, we provide histograms and calibration plots from the ensemble prediction, evaluated on the test dataset.
Test set evaluation metrics -- MSE, MAE, RMSE -- are provided in the main text to quantify the performance of the ensemble model prediction; here we show the data graphically.
In \cref{fig:models_forward_pred}, the plots on the left are histograms showing the relative prediction distribution: the predicted value normalised by the true value. 
In all histograms, the distribution is sharply peaked around \num{1}, indicating reliable prediction of the models.
Note that the $y$-scale is logarithmic.
The $x$-axes of the left-hand plots are truncated to exclude the $0.1\%$-density upper and lower tails to improve visual clarity. 
The plots on the right show 2-d histograms of the predicted values on the $y$-axis against the true values on the $x$-axis. 
All shape parameter predictions show a high degree of correlation with the true values.
There are predictions that lie off the $1$-to-$1$ correlation line (shown in red).
However, the density of these outliers is very low.

$P_{R_\text{out}}$ exhibits a hard boundary at $\sim \SI{1.5}{\metre}$, which corresponds to the maximum value that $R_\text{out}$ can take due to the plasma touching the outer wall.

\clearpage
\section{Coil current shift values}
\label{app:coil_current_shift_values}
This appendix section lists the values of the current shifts in $\bm I_\text{shape}$, in Amp{\`e}res, used for the finite difference calculations in \cref{sec:finite_diff_vs_auto}.
\begin{table}[h!t]
\centering
\begin{tabular}{l|cccccccccc}
\toprule
 & \multicolumn{10}{c}{Coil current shifts ([\SI{}{\ampere}])} \\
 & P4 & P5 & Px & D1 & D2 & D3 & D5 & D6 & D7 & Dp \\
\hline
Small  & 4.3  & 2.0  & 23.3  & 12.7  & 16.3  & 16.3  & 6.3  & 6.3  & 5.7  & 7.3 \\
Medium & 13.0 & 6.0  & 70.0  & 38.0  & 49.0  & 49.0  & 19.0 & 19.0 & 17.0 & 22.0 \\
Large  & 39.0 & 18.0 & 210.0 & 114.0 & 147.0 & 147.0 & 57.0 & 57.0 & 51.0 & 66.0 \\
\bottomrule
\end{tabular}
\caption{Coil current shift values, in Amp{\`e}res, corresponding to the `small', `medium' and `large' shifts used for comparison. 
}
\label{tab:fin_diff_shift_floors}
\end{table}

\clearpage
\section{Shift distributions}
\label{app:all_histograms}
This appendix section contains histograms and for all requested and realised shifts for all the emulated shape parameters. 

\begin{figure}[h!t]
    \centering
    \includegraphics[width=1\linewidth]{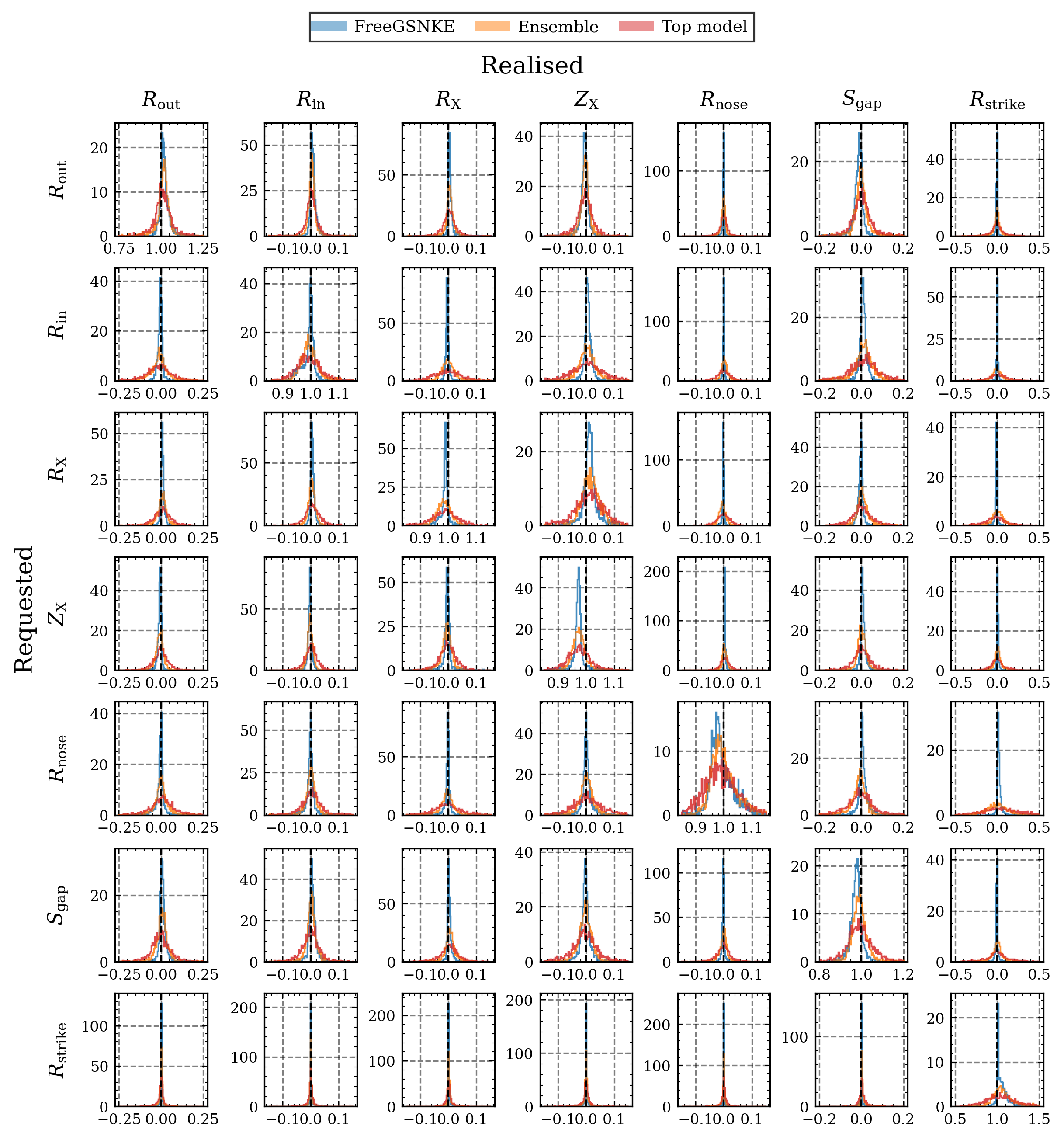}
    \caption{Histograms of realised shifts in \cref{eq:requested_shift}, normalised to the requested shift $\delta \bm P _\text{req}$.
    This plot compares finite-difference GS derivatives (blue) with emulator VCs from the ensemble (orange) and the top model (red) evaluated on early-chain equilibria (MCMC step $< 50$) within the test set.
    This shows that the GS VCs perform best, and that an ensemble of models performs better than the single top model. }
\label{fig:pred_shifts_hists_all_params_FGS_ensemble_single}
\end{figure}

\begin{figure}[h!t]
    \centering
    \includegraphics[width=1\linewidth]{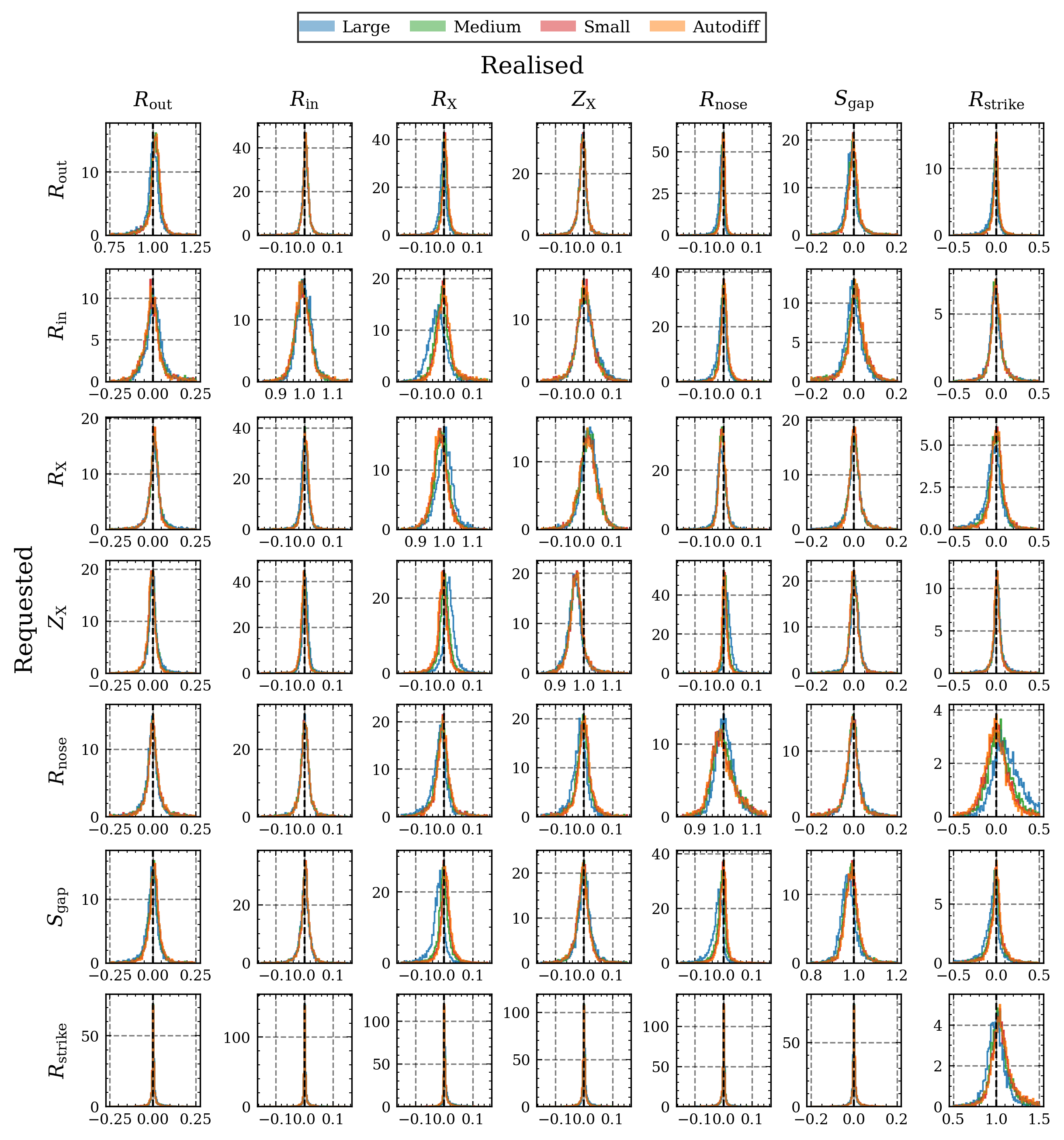}
    \caption{Histograms of realised shifts in \cref{eq:requested_shift}, normalised to the requested shift $\delta \bm P _\text{req}$. This plot compares emulator VCs computed via finite-difference derivatives with different step sizes (blue, green and red) with automatic differentiation (orange) evaluated on early-chain equilibria (MCMC step $< 50$) within the test set.}
    \label{fig:pred_shifts_hists_all_params_finte_vs_auto}
\end{figure}

\Cref{fig:pred_shifts_hists_all_params_FGS_ensemble_single,fig:pred_shifts_hists_all_params_finte_vs_auto} show the distributions of realised shifts, as in \cref{eq:requested_shift}, normalised by the requested shifts, and are the companion plots for \cref{fig:pred_shifts_hists_smallgrid_fgs_ensemble_single,fig:pred_shifts_hists_smallgrid_finite_vs_auto}, respectively.
Here we can see that the majority of the core shape parameters perform well with narrow distributions, while the strike point is consistently wider for all methods of computing the VCs.

\clearpage
\section{Shift tables}
\label{app:all_tables}

The tables below show the means and standard deviations of the realised shifts, corresponding to \cref{app:all_histograms}.
The realised shift values reported in all tables are normalised by the requested shift of \SI{5}{\milli\metre}. Table rows correspond to the requested shift parameter and the columns are the realised shifts.

We show tables for four cases with VCs from the top model with automatic differentiation, the ensemble of models with automatic differentiation, the ensemble of models with finite differences with medium shifts, and the finite difference GS VCs.

\Cref{tab:pred_means_stdev_top_model} shows the results from the top NN model. The majority of the core shape parameters have errors in the few percent range up to about $9\%$, while the realised shifts in $R_\text{strike}$ have generally higher errors with a maximum of $25\%$. 
Comparing this with \cref{tab:pred_means_stdev_ensemble_auto}, where errors are typically up to $5\%$ for all parameters except the strike point, which has errors up to $15\%$, shows that the ensemble of models generally provides more performant VCs.

The finite difference results shown in \cref{tab:pred_means_stdev_finite_medium} are very similar to the automatic differentiation results.

Finally, the results from the finite-difference GS VCs in \cref{tab:pred_means_stdev_FreeGSNKE} show errors on the main shape targets up to $3\%$, with many having sub-percent accuracy. The errors in $R_\text{strike}$ are much closer to the other parameters here, with the largest error being around $5\%$.

\begin{table}[h!t]
\centering
\caption{Means and standard deviations of the realised shift values, as in \cref{eq:requested_shift}, for the single top model, evaluated using automatic differentiation.
Columns correspond to realised shifts and rows to requested shifts.}
\label{tab:pred_means_stdev_top_model}
    \begin{subtable}{\linewidth}
        \centering
        \caption{Means of normalised realised shifts from the automatic-differentiation-based derivatives of the single top model.}
        \begin{tabular}{l|rrrrrrr}
        \toprule
        $\delta \bm P _{req} $ $\backslash$ $\delta \bm P_{GS}$ & $R_\text{out}$ & $R_\text{in}$ & $R_X$ & $Z_X$ & $R_\text{nose}$ & $S_\text{gap}$ & $R_\text{strike}$ \\
        \midrule
        $R_\text{out}$ & 1.0071 & 0.0013 & 0.0026 & -0.0097 & -0.0012 & 0.0008 & -0.0009 \\
        $R_\text{in}$ & 0.0236 & 0.9998 & 0.0053 & -0.0055 & -0.0008 & -0.0259 & 0.0575 \\
        $R_X$ & 0.0025 & 0.0041 & 0.9877 & 0.0093 & -0.0026 & 0.0006 & -0.0639 \\
        $Z_X$ & 0.0016 & 0.0037 & -0.0018 & 0.9729 & 0.0036 & -0.0088 & -0.0121 \\
        $R_\text{nose}$ & 0.0196 & 0.0106 & -0.0017 & 0.0001 & 0.9994 & -0.0300 & 0.1468 \\
        $S_\text{gap}$ & -0.0092 & -0.0049 & -0.0001 & -0.0087 & -0.0025 & 0.9989 & -0.0243 \\
        $R_\text{strike}$ & -0.0106 & -0.0027 & 0.0015 & -0.0033 & -0.0097 & 0.0103 & 1.4378 \\
        \bottomrule
        \end{tabular}
    \end{subtable}

    \vspace{1em} 
    \begin{subtable}{\linewidth}
        \centering
        \caption{Standard deviations of the realised shifts from the automatic-differentiation-based derivatives of the single top model.}
        \begin{tabular}{l|rrrrrrr}
        \toprule
        $\delta \bm P _{req} $ $\backslash$ $\delta \bm P_{GS}$& $R_\text{out}$ & $R_\text{in}$ & $R_\text{X}$ & $Z_\text{X}$ & $R_\text{nose}$ & $S_\text{gap}$ & $R_\text{strike}$ \\
        \midrule
        $R_\text{out}$ & 0.0474 & 0.0188 & 0.0219 & 0.0267 & 0.0133 & 0.0417 & 0.0748 \\
        $R_\text{in}$ & 0.0919 & 0.0450 & 0.0534 & 0.0591 & 0.0258 & 0.0805 & 0.1260 \\
        $R_\text{X}$ & 0.0522 & 0.0256 & 0.0425 & 0.0517 & 0.0238 & 0.0474 & 0.1361 \\
        $Z_\text{X}$ & 0.0445 & 0.0219 & 0.0282 & 0.0392 & 0.0169 & 0.0418 & 0.0848 \\
        $R_\text{nose}$ & 0.0768 & 0.0362 & 0.0508 & 0.0594 & 0.0657 & 0.0675 & 0.2550 \\
        $S_\text{gap}$ & 0.0587 & 0.0290 & 0.0337 & 0.0429 & 0.0244 & 0.0601 & 0.1419 \\
        $R_\text{strike}$ & 0.0214 & 0.0095 & 0.0124 & 0.0139 & 0.0102 & 0.0189 & 0.2008 \\
        \bottomrule
        \end{tabular}
    \end{subtable}
\end{table}

\begin{table}[h!t]
\caption{Means and standard deviations of the realised shift values, as in \cref{eq:requested_shift}, for the NN ensemble, evaluated using automatic differentiation. Columns correspond to realised shifts and rows to requested shifts.}
\label{tab:pred_means_stdev_ensemble_auto}
\centering
    \begin{subtable}{\linewidth}
        \centering
        \caption{Means of normalised realised shifts from the automatic-differentiation-based derivatives of the ensemble of models.}
        \begin{tabular}{l|rrrrrrr}
        \toprule
        $\delta \bm P _{req} $ $\backslash$ $\delta \bm P_{GS}$ & $R_\text{out}$ & $R_\text{in}$ & $R_X$ & $Z_X$ & $R_\text{nose}$ & $S_\text{gap}$ & $R_\text{strike}$ \\
        \midrule
        $R_\text{out}$ & 1.0036 & -0.0004 & 0.0000 & -0.0018 & -0.0020 & 0.0047 & -0.0155 \\
        $R_\text{in}$ & 0.0330 & 1.0202 & 0.0108 & -0.0116 & 0.0016 & -0.0361 & 0.0559 \\
        $R_X$ & 0.0037 & 0.0031 & 0.9842 & 0.0146 & -0.0060 & 0.0041 & -0.1087 \\
        $Z_X$ & -0.0086 & -0.0032 & -0.0069 & 0.9739 & 0.0023 & -0.0005 & -0.0096 \\
        $R_\text{nose}$ & -0.0015 & 0.0170 & -0.0041 & -0.0055 & 1.0093 & -0.0142 & 0.0033 \\
        $S_\text{gap}$ & 0.0008 & -0.0003 & 0.0036 & -0.0040 & -0.0032 & 0.9930 & 0.0577 \\
        $R_\text{strike}$ & 0.0048 & 0.0043 & 0.0050 & -0.0009 & -0.0113 & -0.0015 & 1.4330 \\
        \bottomrule
        \end{tabular}
    \end{subtable}

    \vspace{1em}   
    \begin{subtable}{\linewidth}
        \centering
        \caption{Standard deviations of the realised shifts from the automatic-differentiation-based derivatives of the ensemble of models.}
        \begin{tabular}{l|rrrrrrr}
        \toprule
         $\delta \bm P _{req} $ $\backslash$ $\delta \bm P_{GS}$& $R_\text{out}$ & $R_\text{in}$ & $R_X$ & $Z_X$ & $R_\text{nose}$ & $S_\text{gap}$ & $R_\text{strike}$ \\
        \midrule
        $R_\text{out}$ & 0.0309 & 0.0106 & 0.0116 & 0.0143 & 0.0070 & 0.0252 & 0.0360 \\
        $R_\text{in}$ & 0.0565 & 0.0282 & 0.0258 & 0.0296 & 0.0125 & 0.0479 & 0.0636 \\
        $R_X$ & 0.0265 & 0.0118 & 0.0260 & 0.0309 & 0.0123 & 0.0243 & 0.0773 \\
        $Z_X$ & 0.0222 & 0.0104 & 0.0168 & 0.0217 & 0.0082 & 0.0200 & 0.0412 \\
        $R_\text{nose}$ & 0.0378 & 0.0163 & 0.0240 & 0.0245 & 0.0442 & 0.0348 & 0.1422 \\
        $S_\text{gap}$ & 0.0303 & 0.0145 & 0.0166 & 0.0215 & 0.0124 & 0.0314 & 0.0676 \\
        $R_\text{strike}$ & 0.0097 & 0.0040 & 0.0058 & 0.0061 & 0.0056 & 0.0089 & 0.1243 \\
        \bottomrule
        \end{tabular}
    \end{subtable}
\end{table}

\begin{table}[h!t]
\caption{Means and standard deviations of the realised shift values, as in \cref{eq:requested_shift}, for the NN ensemble, evaluated using finite differences with the medium set of current shifts. Columns correspond to realised shifts and rows to requested shifts. 
}
\label{tab:pred_means_stdev_finite_medium}
\centering
    \begin{subtable}{\linewidth}
        \centering
        \caption{Means of normalised realised shifts from the finite-difference-based derivatives with medium shifts of the ensemble of models.}
       \begin{tabular}{l|rrrrrrr}
        \toprule
        $\delta \bm P _{req} $ $\backslash$ $\delta \bm P_{GS}$ & $R_\text{out}$ & $R_\text{in}$ & $R_X$ & $Z_X$ & $R_\text{nose}$ & $S_\text{gap}$ & $R_\text{strike}$ \\
        \midrule
        $R_\text{out}$ & 1.0007 & -0.0006 & -0.0024 & -0.0020 & -0.0056 & 0.0001 & 0.0293 \\
        $R_\text{in}$ & 0.0294 & 1.0237 & 0.0018 & -0.0118 & -0.0026 & -0.0362 & 0.0011 \\
        $R_X$ & 0.0064 & 0.0018 & 0.9897 & 0.0164 & -0.0073 & 0.0017 & -0.2072 \\
        $Z_X$ & -0.0041 & -0.0010 & 0.0016 & 0.9719 & 0.0072 & -0.0015 & 0.0458 \\
        $R_\text{nose}$ & 0.0014 & 0.0047 & -0.0073 & -0.0146 & 1.0124 & -0.0179 & 0.1357 \\
        $S_\text{gap}$ & -0.0044 & -0.0003 & -0.0058 & -0.0025 & -0.0128 & 0.9874 & -0.0841 \\
        $R_\text{strike}$ & 0.0041 & 0.0039 & 0.0074 & 0.0005 & -0.0153 & -0.0031 & 1.4841 \\
        \bottomrule
        \end{tabular}

    \end{subtable}

    \vspace{1em} 
    \begin{subtable}{\linewidth}
        \centering
        \caption{Standard deviations of the realised shifts from the finite-difference-based derivatives with medium shifts of the ensemble of models.}
        \begin{tabular}{l|rrrrrrr}
        \toprule
        $\delta \bm P _{req} $ $\backslash$ $\delta \bm P_{GS}$ & $R_\text{out}$ & $R_\text{in}$ & $R_X$ & $Z_X$ & $R_\text{nose}$ & $S_\text{gap}$ & $R_\text{strike}$ \\
        \midrule
        $R_\text{out}$ & 0.0315 & 0.0105 & 0.0115 & 0.0142 & 0.0074 & 0.0251 & 0.0375 \\
        $R_\text{in}$ & 0.0539 & 0.0277 & 0.0261 & 0.0301 & 0.0129 & 0.0448 & 0.0651 \\
        $R_X$ & 0.0280 & 0.0117 & 0.0260 & 0.0297 & 0.0128 & 0.0256 & 0.0800 \\
        $Z_X$ & 0.0230 & 0.0104 & 0.0168 & 0.0223 & 0.0087 & 0.0206 & 0.0422 \\
        $R_\text{nose}$ & 0.0380 & 0.0163 & 0.0247 & 0.0245 & 0.0405 & 0.0343 & 0.1521 \\
        $S_\text{gap}$ & 0.0312 & 0.0147 & 0.0167 & 0.0219 & 0.0140 & 0.0328 & 0.0699 \\
        $R_\text{strike}$ & 0.0098 & 0.0041 & 0.0069 & 0.0062 & 0.0056 & 0.0087 & 0.1151 \\
        \bottomrule
        \end{tabular}
    \end{subtable}
\end{table}

\begin{table}[h!t]
\centering
\caption{Means and standard deviations of the realised shift values, as in \cref{eq:requested_shift}, for finite-difference-based GS VCs computed with FreeGSNKE. Columns correspond to realised shifts and rows to requested shifts.}
\label{tab:pred_means_stdev_FreeGSNKE}
    \begin{subtable}{\linewidth}
        \centering
        \caption{Means of normalised realised shifts from the finite-difference-based GS VCs computed with FreeGSNKE.}
        \label{tab:FreeGSNKE_normalised_mean}
        \begin{tabular}{l|rrrrrrr}
        \toprule
        $\delta \bm P _{req} $ $\backslash$ $\delta \bm P_{GS}$ & $R_\text{out}$ & $R_\text{in}$ & $R_X$ & $Z_X$ & $R_\text{nose}$ & $S_\text{gap}$ & $R_\text{strike}$ \\
        \midrule
        $R_\text{out}$ & 1.0144 & 0.0105 & 0.0036 & -0.0065 & -0.0025 & -0.0194 & -0.0103 \\
        $R_\text{in}$ & -0.0080 & 0.9596 & -0.0041 & 0.0066 & -0.0018 & 0.0010 & 0.1123 \\
        $R_X$ & 0.0121 & 0.0083 & 0.9838 & 0.0083 & -0.0027 & -0.0108 & -0.1965 \\
        $Z_X$ & -0.0099 & -0.0032 & -0.0052 & 0.9654 & 0.0034 & 0.0050 & -0.0236 \\
        $R_\text{nose}$ & -0.0043 & -0.0006 & 0.0070 & -0.0126 & 0.9918 & 0.0005 & 0.2470 \\
        $S_\text{gap}$ & 0.0071 & 0.0043 & -0.0008 & 0.0086 & -0.0030 & 0.9734 & -0.0619 \\
        $R_\text{strike}$ & 0.0027 & 0.0030 & -0.0011 & -0.0001 & 0.0005 & -0.0043 & 1.2467 \\
        \bottomrule
        \end{tabular}
    \end{subtable}

    \vspace{1em} 
    \begin{subtable}{\linewidth}
        \centering
        \caption{Standard deviations of normalised realised shifts from the finite-difference-based GS VCs computed with FreeGSNKE.}
        \label{tab:FreeGSNKE_normalised_std_pct}
        \begin{tabular}{l|rrrrrrr}
        \toprule
        $\delta \bm P _{req} $ $\backslash$ $\delta \bm P_{GS}$ & $R_\text{out}$ & $R_\text{in}$ & $R_X$ & $Z_X$ & $R_\text{nose}$ & $S_\text{gap}$ & $R_\text{strike}$ \\
        \midrule
        $R_\text{out}$ & 0.0213 & 0.0086 & 0.0057 & 0.0113 & 0.0021 & 0.0171 & 0.0065 \\
        $R_\text{in}$ & 0.0118 & 0.0210 & 0.0049 & 0.0113 & 0.0024 & 0.0139 & 0.0063 \\
        $R_X$ & 0.0087 & 0.0055 & 0.0116 & 0.0189 & 0.0023 & 0.0096 & 0.0098 \\
        $Z_X$ & 0.0086 & 0.0059 & 0.0076 & 0.0109 & 0.0019 & 0.0092 & 0.0057 \\
        $R_\text{nose}$ & 0.0146 & 0.0093 & 0.0061 & 0.0123 & 0.0385 & 0.0157 & 0.0172 \\
        $S_\text{gap}$ & 0.0150 & 0.0099 & 0.0060 & 0.0140 & 0.0033 & 0.0223 & 0.0092 \\
        $R_\text{strike}$ & 0.0033 & 0.0022 & 0.0018 & 0.0025 & 0.0009 & 0.0036 & 0.0553 \\
        \bottomrule
        \end{tabular}
    \end{subtable}
\end{table}

\clearpage
\bibliographystyle{unsrt} 
\bibliography{refs}

@article{freegsnke,
    author = {Amorisco, N. C. and Agnello, A. and Holt, G. and Mars, M. and Buchanan, J. and Pamela, S.},
    title = {FreeGSNKE: A Python-based dynamic free-boundary toroidal plasma equilibrium solver},
    journal = {Physics of Plasmas},
    volume = {31},
    number = {4},
    pages = {042517},
    year = {2024},
    month = {04},
    issn = {1070-664X},
    doi = {10.1063/5.0188467},
    url = {https://doi.org/10.1063/5.0188467},
    eprint = {https://pubs.aip.org/aip/pop/article-pdf/doi/10.1063/5.0188467/19912488/042517_1_5.0188467.pdf},
}

@INPROCEEDINGS{Emulated_VC_IEEE,
  author={Cavestany, P. and Ross, A. and Agnello, A. and Garrod, A. and Amorisco, N. C. and Holt, G. K. and Pentland, K. and Buchanan, J.},
  booktitle={2025 IEEE Conference on Control Technology and Applications (CCTA)}, 
  title={Real-Time Applicability of Emulated Virtual Circuits for Tokamak Plasma Shape Control}, 
  year={2025},
  volume={},
  number={},
  pages={826-831},
  doi={10.1109/CCTA53793.2025.11151371}}

@article{FPDT,
  title={The FreeGSNKE Pulse Design Tool ({FPDT}): a computational framework for evolutive plasma scenario and control design},
  author={Pentland, K. and Amorisco, N. C. and Ross, A. and Cavestany, P. and Nunn, T. and Agnello, A. and Holt, G. K. and McArdle, G. and Vincent, C. and Buchanan, J. and Pamela, S. J. P.},
  journal={Plasma Physics and Controlled Fusion},
  volume={68},
  number={7},
  pages={075035},
  year={2026},
  publisher={IOP Publishing}
}

@article{dynamic_val_paper,
  title={Real-time virtual circuits for plasma shape control via neural network surrogates: dynamic validation in closed-loop simulations},
  author={Pentland, K. and Ross, A. and Amorisco, N. C. and Cavestany, P. and Nunn, T. and Agnello, A. and Holt, G. K. and Vincent, C.},
  journal={arXiv preprint arXiv:2604.00781},
  year={2026}
}

@article{HORNIK1989,
	author = {K. Hornik and M. Stinchcombe and H. White},
	doi = {https://doi.org/10.1016/0893-6080(89)90020-8},
	issn = {0893-6080},
	journal = {Neural Networks},
	number = {5},
	pages = {359-366},
	title = {Multilayer feedforward networks are universal approximators},
	url = {https://www.sciencedirect.com/science/article/pii/0893608089900208},
	volume = {2},
	year = {1989}}

@article{HORNIK1990,
	author = {K. Hornik and M. Stinchcombe and H. White},
	doi = {https://doi.org/10.1016/0893-6080(90)90005-6},
	issn = {0893-6080},
	journal = {Neural Networks},
	number = {5},
	pages = {551-560},
	title = {Universal approximation of an unknown mapping and its derivatives using multilayer feedforward networks},
	url = {https://www.sciencedirect.com/science/article/pii/0893608090900056},
	volume = {3},
	year = {1990}}

@article{HORNIK1991,
	author = {K. Hornik},
	doi = {https://doi.org/10.1016/0893-6080(91)90009-T},
	issn = {0893-6080},
	journal = {Neural Networks},
	number = {2},
	pages = {251-257},
	title = {Approximation capabilities of multilayer feedforward networks},
	url = {https://www.sciencedirect.com/science/article/pii/089360809190009T},
	volume = {4},
	year = {1991}}

@article{LESHNO1993,
	author = {M. Leshno and V. Y. Lin and A. Pinkus and S. Schocken},
	doi = {https://doi.org/10.1016/S0893-6080(05)80131-5},
	issn = {0893-6080},
	journal = {Neural Networks},
	number = {6},
	pages = {861-867},
	title = {Multilayer feedforward networks with a nonpolynomial activation function can approximate any function},
	url = {https://www.sciencedirect.com/science/article/pii/S0893608005801315},
	volume = {6},
	year = {1993}}

@article{Cybenko1989,
	author = {Cybenko, G. },
	date = {1989/12/01},
	doi = {10.1007/BF02551274},
	id = {Cybenko1989},
	isbn = {1435-568X},
	journal = {Mathematics of Control, Signals and Systems},
	number = {4},
	pages = {303--314},
	title = {Approximation by superpositions of a sigmoidal function},
	url = {https://doi.org/10.1007/BF02551274},
	volume = {2},
	year = {1989}}

@article{Agnello24,
    author = {Agnello, A. and Amorisco, N. C. and Keats, A. and Holt, G. K. and Buchanan, J. and Pamela, S. and Vincent, C. and McArdle, G.},
    title = {Emulation techniques for scenario and classical control design of tokamak plasmas},
    journal = {Physics of Plasmas},
    volume = {31},
    number = {4},
    pages = {043901},
    year = {2024},
    month = {04},
    issn = {1070-664X},
    doi = {10.1063/5.0187822},
    url = {https://doi.org/10.1063/5.0187822},
    eprint = {https://pubs.aip.org/aip/pop/article-pdf/doi/10.1063/5.0187822/19864754/043901_1_5.0187822.pdf},
}

@article{Lao_1985,
	author = {Lao, L.L. and St. John, H. and Stambaugh, R.D. and Kellman, A.G. and Pfeiffer, W.},
	doi = {10.1088/0029-5515/25/11/007},
	journal = {Nuclear Fusion},
	month = {nov},
	number = {11},
	pages = {1611},
	title = {Reconstruction of current profile parameters and plasma shapes in tokamaks},
	url = {https://doi.org/10.1088/0029-5515/25/11/007},
	volume = {25},
	year = {1985}}

@article{wai2026tutorialinversionbasedshapecontrol,
  title={A tutorial on inversion-based shape control with design application to {NSTX-U}},
  author={Wai, J. T. and Boyer, M. D. and Battaglia, D. J. and Carpanese, F. and Felici, F. and Wehner, W. P. and Welander, A. S. and Kolemen, E.},
  journal={arXiv preprint arXiv:2602.18667},
  year={2026}
}

@article{DBLP:journals/corr/CzarneckiOJSP17,
  author       = {W. M. Czarnecki and
                  S. Osindero and
                  M. Jaderberg and
                  G. Swirszcz and
                  R. Pascanu},
  title        = {Sobolev Training for Neural Networks},
  journal      = {CoRR},
  volume       = {abs/1706.04859},
  year         = {2017},
  url          = {http://arxiv.org/abs/1706.04859},
  eprinttype   = {arXiv},
  eprint       = {1706.04859},
  bibsource    = {dblp computer science bibliography, https://dblp.org}
}

@article{bishop1995,
  author  = {Bishop, C. M. and Haynes, P. S. and Smith, M. E. U. and
             Todd, T. N. and Trotman, D. L.},
  title   = {Real-Time Control of a Tokamak Plasma Using Neural Networks},
  journal = {Neural Computation},
  volume  = {7},
  number  = {1},
  pages   = {206--217},
  year    = {1995},
  doi     = {10.1162/neco.1995.7.1.206}
}

@article{wai2022,
  author  = {Wai, J. T. and Boyer, M. D. and Kolemen, E.},
  title   = {Neural net modeling of equilibria in {NSTX-U}},
  journal = {Nuclear Fusion},
  volume  = {62},
  number  = {8},
  pages   = {086042},
  year    = {2022},
  doi     = {10.1088/1741-4326/ac77e6}
}

@article{lao2022,
  author  = {Lao, L. L. and Kruger, S. and Akcay, C. and Balaprakash, P. and
             Bechtel, T. A. and Howell, E. and Koo, J. and Leddy, J. and
             Leinhauser, M. and Liu, Y. Q. and Madireddy, S. and
             McClenaghan, J. and Orozco, D. and Pankin, A. and Schissel, D.
             and Smith, S. and Sun, X. and Williams, S.},
  title   = {Application of machine learning and artificial intelligence to
             extend {EFIT} equilibrium reconstruction},
  journal = {Plasma Physics and Controlled Fusion},
  volume  = {64},
  number  = {7},
  pages   = {074001},
  year    = {2022},
  doi     = {10.1088/1361-6587/ac6fff}
}

@article{rui2025,
  author  = {W. Rui and Y. Wang and H. Song and Z. Huang and Z. Luo and Y. Huang and Z. Liu and K. Wu and J. Huang and B. Xiao},
  title   = {Adaptive vertical position control system based on neural networks},
  journal = {Nuclear Fusion},
  volume  = {66},
  number  = {2},
  pages   = {026012},
  year    = {2025},
  doi     = {10.1088/1741-4326/ae2693}
}

@article{rasouli2013,
  author  = {Rasouli, H. and Rasouli, C. and Koohi, A.},
  title   = {Identification and control of plasma vertical position using
             neural network in {Damavand} tokamak},
  journal = {Review of Scientific Instruments},
  volume  = {84},
  number  = {2},
  pages   = {023504},
  year    = {2013},
  doi     = {10.1063/1.4791925}
}

@inproceedings{DeTommasi2022,
  author    = {De Tommasi, G. and Dubbioso, S. and Huang, Y. and
               Luo, Z.-P. and Mele, A. and Xiao, B. J.},
  title     = {A {RL}-based Vertical Stabilization System for the {EAST}
               tokamak},
  booktitle = {2022 American Control Conference (ACC)},
  address   = {Atlanta, GA, USA},
  pages     = {5328--5333},
  year      = {2022},
  doi       = {10.23919/ACC53348.2022.9867499}
}

@article{degrave2022,
  author  = {Degrave, J. and Felici, F. and Buchli, J. and
             Neunert, M. and Tracey, B. and Carpanese, F. and
             Ewalds, T. and Hafner, R. and Abdolmaleki, A. and
             {de las Casas}, D. and Donner, C. and Fritz, L. and
             Galperti, C. and Huber, A. and Keeling, J. and
             Tsimpoukelli, M. and Kay, J. and Merle, A. and
             Moret, J.-M. and Noury, S. and Pesamosca, F. and
             Pfau, D. and Sauter, O. and Sommariva, C. and
             Coda, S. and Duval, B. and Fasoli, A. and
             Kohli, P. and Kavukcuoglu, K. and Hassabis, D. and
             Riedmiller, M.},
  title   = {Magnetic control of tokamak plasmas through deep reinforcement
             learning},
  journal = {Nature},
  volume  = {602},
  pages   = {414--419},
  year    = {2022},
  doi     = {10.1038/s41586-021-04301-9}
}

@article{seo2024,
  author  = {Seo, J. and Kim, S. and Jalalvand, A. and
             Conlin, R. and Rothstein, A. and Abbate, J. and
             Erickson, K. and Wai, J. and Shousha, R. and
             Kolemen, E.},
  title   = {Avoiding fusion plasma tearing instability with deep
             reinforcement learning},
  journal = {Nature},
  volume  = {626},
  number  = {8000},
  pages   = {746--751},
  year    = {2024},
  doi     = {10.1038/s41586-024-07024-9}
}

@article{tracey2024,
  author  = {Tracey, B. D. and Michi, A. and Chervonyi, Y. and
             Davies, I. and Paduraru, C. and Lazic, N. and
             Felici, F. and Ewalds, T. and Donner, C. and
             Galperti, C. and Buchli, J. and Neunert, M. and
             Huber, A. and Evens, J. and Kurylowicz, P. and
             Mankowitz, D. J. and Riedmiller, M. and
             {The TCV Team}},
  title   = {Towards practical reinforcement learning for tokamak magnetic
             control},
  journal = {Fusion Engineering and Design},
  volume  = {200},
  pages   = {114161},
  year    = {2024},
  doi     = {10.1016/j.fusengdes.2024.114161}
}

@article{kerboua2024,
  author  = {Kerboua-Benlarbi, S. and Nouailletas, R. and
             Faugeras, B. and Nardon, E. and Moreau, P.},
  title   = {Magnetic control of {WEST} plasmas through deep reinforcement
             learning},
  journal = {{IEEE} Transactions on Plasma Science},
  volume  = {52},
  pages   = {3698--3703},
  year    = {2024},
  doi     = {10.1109/TPS.2024.3377811}
}

@article{Anand_2024,
	author = {Anand, H. and Wehner, W. and Eldon, D. and Welander, A. and Xing, Z. and Lvovskiy, A. and Barr, J. and Cho, E. and Sammuli, B. and Humphreys, D. and Eidietis, N. and Leonard, A. and Kochan, M. and Vincent, C. and McArdle, G. and Cunningham, G. and Thornton, A. and Harrison, J. and Soukhanovskii, V. and Lovell, J.},
	doi = {10.1088/1741-4326/ad5c80},
	journal = {Nuclear Fusion},
	month = {jul},
	number = {8},
	pages = {086051},
	publisher = {IOP Publishing},
	title = {Real-time plasma equilibrium reconstruction and shape control for the {MAST} {Upgrade} tokamak},
	url = {https://doi.org/10.1088/1741-4326/ad5c80},
	volume = {64},
	year = {2024}}

@inproceedings{bishop1992,
	address = {London},
	author = {Bishop, C. and Cox, P. and Haynes, P. and Roach, C. and Smith, M. and Todd, T. and Trotman, D.},
	booktitle = {Neural Network Applications},
	editor = {Taylor, J. G.},
	isbn = {978-1-4471-2003-2},
	pages = {114--128},
	publisher = {Springer London},
	title = {A Neural Network Approach to Tokamak Equilibrium Control},
	year = {1992}}

@book{RL_book_SuttonB98,
  author       = {R. S. Sutton and
                  A. G. Barto},
  title        = {Reinforcement learning - an introduction},
  series       = {Adaptive computation and machine learning},
  publisher    = {{MIT} Press},
  year         = {1998},
  url          = {http://www.incompleteideas.net/book/first/the-book.html},
  isbn         = {978-0-262-19398-6},
  bibsource    = {dblp computer science bibliography, https://dblp.org}
}

@inproceedings{kochan2023,
  author    = {Kochan, M. and Anand, H. and Lvovskiy, A. and Ryan, P. and
               Verhaegh, K. and Wijkamp, T. and Kirk, A. and McArdle, G.},
  title     = {Real-time plasma shape reconstruction on {MAST} {Upgrade}
               based on local expansion},
  booktitle = {30th {IEEE} Symposium on Fusion Engineering ({SOFE})},
  address   = {Oxford, UK},
  year      = {2023},
  note      = {Conference presentation, 9--13 July 2023}
}

@inproceedings{akiba2019optuna,
  title={{O}ptuna: A Next-Generation Hyperparameter Optimization Framework},
  author={Akiba, T. and Sano, S. and Yanase, T. and Ohta, T. and Koyama, M.},
  booktitle={The 25th ACM SIGKDD International Conference on Knowledge Discovery \& Data Mining},
  pages={2623--2631},
  year={2019}
}

@misc{RTVC,
      title={Real-time virtual circuits for plasma shape control via neural network emulators: integration and testing in the {MAST-U} {PCS}}, 
      author={M. J. Marshall and E. Jones and G. J. McArdle and A. Ross and K. Pentland and N. C. Amorisco and C. Vincent and M. Kochan and C. Hogben and G. Jones and A. Stephen and G. K. Holt and A. Agnello},
      year={2026},
      eprint={2608.26216},
      archivePrefix={arXiv},
      primaryClass={physics.plasm-ph},
      url={https://arxiv.org/abs/2608.26216}, 
}

@article{PROKHOROV2020857,
	author = {A. A. Prokhorov and Y. V. Mitrishkin and P. S. Korenev and M. I. Patrov},
	doi = {https://doi.org/10.1016/j.ifacol.2020.12.843},
	issn = {2405-8963},
	journal = {IFAC-PapersOnLine},
	note = {21st IFAC World Congress},
	number = {2},
	pages = {857-862},
	title = {The plasma shape control system in the tokamak with the artificial neural network as a plasma equilibrium reconstruction algorithm},
	url = {https://www.sciencedirect.com/science/article/pii/S2405896320311678},
	volume = {53},
	year = {2020}}

@article{MCARDLE2020111764,
	author = {G. McArdle and L. Pangione and M. Kochan},
	doi = {https://doi.org/10.1016/j.fusengdes.2020.111764},
	issn = {0920-3796},
	journal = {Fusion Engineering and Design},
	pages = {111764},
	title = {The {MAST} {Upgrade} plasma control system},
	url = {https://www.sciencedirect.com/science/article/pii/S0920379620303124},
	volume = {159},
	year = {2020}}

@ARTICLE{2026NucFu..66b6012R,
       author = {{Rui}, W. and {Wang}, Y. and {Song}, H. and {Huang}, Z. and {Luo}, Z. and {Huang}, Y. and {Liu}, Z. and {Wu}, K. and {Huang}, J. and {Xiao}, B.},
        title = "{Adaptive vertical position control system based on neural networks}",
      journal = {Nuclear Fusion},
         year = 2026,
        month = feb,
       volume = {66},
       number = {2},
          eid = {026012},
        pages = {026012},
          doi = {10.1088/1741-4326/ae2693},
       adsurl = {https://ui.adsabs.harvard.edu/abs/2026NucFu..66b6012R}
}

@book{ariola2008magnetic,
  title={Magnetic Control of Tokamak Plasmas},
  author={Ariola, M. and Pironti, A.},
  isbn={9781848003248},
  lccn={2008927412},
  series={Advances in Industrial Control},
  url={https://books.google.co.in/books?id=lC1UigdNwdYC},
  year={2008},
  publisher={Springer London}
}

@INPROCEEDINGS{Walker_tok_plas_ctrl,
  author={Walker, M. L. and De Vries, P. and Felici, F. and Schuster, E.},
  booktitle={2020 American Control Conference (ACC)}, 
  title={Introduction to Tokamak Plasma Control}, 
  year={2020},
  volume={},
  number={},
  pages={2901-2918},
  doi={10.23919/ACC45564.2020.9147561}}

@ARTICLE{Orozoco_NN_RT_disruption,
  author={Orozco, D. and Sammuli, B. and Barr, J. and Wehner, W. and Humphreys, D.},
  journal={IEEE Transactions on Plasma Science}, 
  title={Neural Network-Based Confinement Mode Prediction for Real-Time Disruption Avoidance}, 
  year={2022},
  volume={50},
  number={11},
  pages={4157-4164},
  doi={10.1109/TPS.2022.3198596}}

@InProceedings{Char_offline_RL_tok_ctrl,
  title = 	 {Offline Model-Based Reinforcement Learning for Tokamak Control},
  author =       {Char, I. and Abbate, J. and Bardoczi, L. and Boyer, M. and Chung, Y. and Conlin, R. and Erickson, K. and Mehta, V. and Richner, N. and Kolemen, E. and Schneider, J.},
  booktitle = 	 {Proceedings of The 5th Annual Learning for Dynamics and Control Conference},
  pages = 	 {1357--1372},
  year = 	 {2023},
  editor = 	 {Matni, Nikolai and Morari, Manfred and Pappas, George J.},
  volume = 	 {211},
  series = 	 {Proceedings of Machine Learning Research},
  month = 	 {15--16 Jun},
  publisher =    {PMLR},
  url = 	 {https://proceedings.mlr.press/v211/char23a.html}
}

@article{Tang-Deep_learning_disruption_prediction,
author = {Tang, W. and Dong, G. and Barr, J. and Erickson, K. and Conlin, R. and Boyer, D. and Kates-Harbeck, J. and Felker, K. and Rea, C. and Logan, N. and Svyatkovskiy, A. and Feibush, E. and Abbatte, J. and Clement, M. and Grierson, B. and Nazikian, R. and Lin, Z. and Eldon, D. and Moser, A. and Maslov, M.},
title = {Implementation of AI/DEEP learning disruption predictor into a plasma control system},
journal = {Contributions to Plasma Physics},
volume = {63},
number = {5-6},
pages = {e202200095},
doi = {https://doi.org/10.1002/ctpp.202200095},
url = {https://onlinelibrary.wiley.com/doi/abs/10.1002/ctpp.202200095},
eprint = {https://onlinelibrary.wiley.com/doi/pdf/10.1002/ctpp.202200095},
year = {2023}
}

@article{pentland2024,
	title = {Validation of the static forward {Grad–Shafranov} equilibrium solvers in {FreeGSNKE} and {Fiesta} using {EFIT}++ reconstructions from {MAST-U}},
	doi = {10.1088/1402-4896/ada192},
	journal = {Physica Scripta},
    volume = {100},
    pages = {025608},
    number = {2},
	author = {Pentland, K. and Amorisco, N. C. and El-Zobaidi, O. and Etches, S. and Agnello, A. and Holt, G. K. and Ross, A. and Vincent, C. and Buchanan, J. and Pamela, S. and {McArdle}, G. and Kogan, L. and Cunningham, G.},
	year = {2024},
}

@article{Mele_2025,
	author = {Mele, A. and Tenaglia, A. and Felici, F. and Galperti, C. and Carnevale, D. and Coda, S. and Merle, A. and Pironti, A. and Sauter, O. and team, the TCV and Exploitation team, the Eurofusion Tokamak},
	doi = {10.1088/1361-6587/addeee},
	journal = {Plasma Physics and Controlled Fusion},
	month = {jun},
	number = {6},
	pages = {065035},
	publisher = {IOP Publishing},
	title = {Design and implementation of a model-based hierarchical architecture for plasma shape control in the {TCV} tokamak},
	url = {https://doi.org/10.1088/1361-6587/addeee},
	volume = {67},
	year = {2025}}

@article{Harrison_2024_SuperX,
  author  = {Harrison, J. R. and Bowman, C. and Clark, J. G. and Kirk, A. and
             Lovell, J. and Patel, B. S. and Ryan, P. and Scannell, R. and
             Thornton, A. J. and Verhaegh, K.},
  title   = {Benefits of the {Super-X} divertor configuration for scenario
             integration on {MAST Upgrade}},
  journal = {Plasma Physics and Controlled Fusion},
  volume  = {66},
  number  = {6},
  pages   = {065019},
  year    = {2024},
  doi     = {10.1088/1361-6587/ad4058}
}

@article{bonotto2024planet,
  author = {Bonotto, M. and Abate, D. and Pigatto, L.},
  title = {Reconstruction of plasma equilibrium and separatrix using convolutional physics-informed neural operator},
  journal = {Fusion Engineering and Design},
  volume = {200},
  pages = {114193},
  year = {2024},
  doi = {10.1016/j.fusengdes.2024.114193},
  url = {https://doi.org/10.1016/j.fusengdes.2024.114193},
}

@article{zheng2024efitnn,
  author = {Zheng, G. H. and Yang, Z. Y. and Liu, S. F. and Ma, R. and Gong, X. W. and Wang, A. and Wang, S. and Zhong, W. L.},
  title = {Real-time equilibrium reconstruction by multi-task learning neural network based on {HL-3} tokamak},
  journal = {Nuclear Fusion},
  volume = {64},
  number = {12},
  pages = {126041},
  year = {2024},
  doi = {10.1088/1741-4326/ad8014},
  url = {https://doi.org/10.1088/1741-4326/ad8014},
}

@article{zheng2025efitmini,
  author = {Zheng, G. H. and Liu, S. F. and Xie, H. S. and Gu, X. and Chen, Z. Y. and Lun, X. C. and Liu, Y. and Li, J. and Guo, D. and Tao, R. Y. and Zhao, H. Y. and Zhang, Y. P. and Sun, T. T. and Xu, Y. N. and Hu, Y. J. and Yang, Z. Y.},
  title = {{EFIT-mini}: an embedded, multi-task neural network-driven equilibrium inversion algorithm},
  journal = {Nuclear Fusion},
  volume = {65},
  number = {10},
  pages = {106008},
  year = {2025},
  doi = {10.1088/1741-4326/adff94},
  url = {https://doi.org/10.1088/1741-4326/adff94},
}

@article{grandin2026surrogate,
  author = {Grandin, M. and Mele, A. and Galperti, C. and Gonzales Castineiras, D. and Hei{\ss}, C. and Cenedese, A. and {TCV team} and {EUROfusion Tokamak Exploitation team}},
  title = {Experimental validation of a fast control-oriented, physics-informed surrogate model for plasma equilibrium reconstruction in the {TCV} tokamak},
  journal = {arXiv preprint arXiv:2606.09487},
  year = {2026},
  doi = {10.48550/arXiv.2606.09487},
  url = {https://arxiv.org/abs/2606.09487v2},
}

@article{fiorenza2026caronte,
  author = {Fiorenza, F. and Dubbioso, S. and De Tommasi, G. and Pironti, A.},
  title = {{CARONTE}: A Physics-Informed Extreme Learning Machine-Based Algorithm for Plasma Boundary Reconstruction in Magnetically Confined Fusion Devices},
  journal = {IEEE Access},
  volume = {14},
  pages = {105705},
  year = {2026},
  doi = {10.1109/access.2026.3712498},
  url = {https://doi.org/10.1109/access.2026.3712498},
}

@article{ariola2005xsc,
  author = {Ariola, M. and Pironti, A.},
  title = {Plasma shape control for the {JET} tokamak: an optimal output regulation approach},
  journal = {IEEE Control Systems Magazine},
  volume = {25},
  number = {5},
  pages = {65},
  year = {2005},
  doi = {10.1109/mcs.2005.1512796},
  url = {https://doi.org/10.1109/mcs.2005.1512796},
}

@article{ambrosino2008xsc,
  author = {Ambrosino, G. and Ariola, M. and Pironti, A. and Sartori, F.},
  title = {Design and Implementation of an Output Regulation Controller for the {JET} Tokamak},
  journal = {IEEE Transactions on Control Systems Technology},
  volume = {16},
  number = {6},
  pages = {1101},
  year = {2008},
  doi = {10.1109/tcst.2008.917216},
  url = {https://doi.org/10.1109/tcst.2008.917216},
}

@article{ambrosino2007steady,
  author = {Ambrosino, G. and Ariola, M. and Pironti, A.},
  title = {Optimal steady-state control for linear non-right-invertible systems},
  journal = {IET Control Theory \& Applications},
  volume = {1},
  number = {3},
  pages = {604},
  year = {2007},
  doi = {10.1049/iet-cta:20060237},
  url = {https://doi.org/10.1049/iet-cta:20060237},
}

@article{ariola2008flux,
  author = {Ariola, M. and De Tommasi, G. and Mazon, D. and Moreau, D. and Piccolo, F. and Pironti, A. and Sartori, F. and Zabeo, L. and {JET-EFDA Contributors}},
  title = {Integrated Plasma Shape and Boundary Flux Control on {JET} Tokamak},
  journal = {Fusion Science and Technology},
  volume = {53},
  number = {3},
  pages = {789},
  year = {2008},
  doi = {10.13182/fst08-a1735},
  url = {https://doi.org/10.13182/fst08-a1735},
}

\end{document}